\documentclass[nofootinbib,nobibnotes,groupaddress,preprintnumbers,aps,pre,twocolumn]{revtex4}
\usepackage{tipa}
\usepackage{bm,epsfig,mathrsfs,amsmath,amssymb,graphicx}
\usepackage{epsfig,amsmath,graphicx,amssymb}
\usepackage{graphicx}
\usepackage{dcolumn}
\usepackage{bm}
\usepackage{float} 
\usepackage{booktabs}%
\newcommand{\PreserveBackslash}[1]{\let\temp=\\#1\let\\=\temp}
\newcolumntype{C}[1]{>{\PreserveBackslash\centering}p{#1}}
\newcolumntype{R}[1]{>{\PreserveBackslash\raggedleft}p{#1}}
\newcolumntype{L}[1]{>{\PreserveBackslash\raggedright}p{#1}}
\usepackage[colorlinks,citecolor=blue,linkcolor=red]{hyperref}
\usepackage[toc,title,titletoc,header]{appendix}

\date{\today}

\begin{document}

\title{Finite-time quantum Otto engine with a squeezed thermal bath: Role of quantum coherence and squeezing in the performance and fluctuations}
\author{Yang Xiao$^{1}$}
\author{Dehua Liu$^1$}
\author{Jizhou He$^{1}$}
\author{Wu-Ming Liu$^{2}$}\email{wmliu@iphy.ac.cn}
\author{Jianhui Wang$^{1,3}$}\email{wangjianhui@ncu.edu.cn}

\affiliation { $^1\,$ Department of Physics, Nanchang University,
Nanchang, 330031, China\\ 
$^2\,$ 
Beijing National Laboratory for Condensed Matter Physics, Institute of Physics, Chinese Academy of Sciences, Beijing 100190, China\\
$^3\,$ State Key Laboratory
of Surface Physics and Department of Physics, Fudan University,
Shanghai 200433, China}
\begin{abstract}
We consider a finite-time quantum Otto heat engine that consists of two  isochoric (thermal-contact) process, where the system is alternatively coupled to a hot squeezed and a cold thermal reservoir,  and  two unitary driven strokes, where the system is isolated from these two baths and its von Neumann entropy keeps constant. Both quantum inner friction and coherence are generated along the driven stroke and coherence cannot be fully erased after the finite-time hot isochore. 
 Using full counting statistics, we present the probability
distribution functions  of heat injection and total work per cycle, which are dependent
on the time duration along each process. With these, we derive  the analytical
expressions for the thermodynamic quantities of the two-level heat engine, such as total work, thermodynamic efficiency, entropy production, and work fluctuations,  in which effects of coherence, squeezing, inner friction and finite-time heat exchange are included. We then numerically
determine the thermodynamic quantities and the fluctuations using the parameters employed in the experimental implementation. Our results clarify the role of coherence and squeezing in the performance and fluctuations in the quantum Otto engines.

PACS number(s):05.70.Ln

\end{abstract}

\maketitle
\date{\today}
\section{Introduction}
Heat engines  produce useful work by
consuming thermal energy.  A great effort has been devoted to the extension from  macroscopic engines to quantum heat engines \cite{Kos14, Nie18, Scu11}. Practically all heat engines should run  far from the ideal maximum efficiency at which the power becomes vanishing; the working system proceeding in irreversible processes can not reach Gibbs thermal state due to finite-time operation.   Therefore, quantum heat engines provide  excellent platforms for understanding nonequilibrium thermodynamic and nonequilibrium statistical  implications of open quantum systems.

 Towards the ever-smaller scale of thermal machines,    quantum   effects increasingly manifest themselves,  such as quantum coherence\cite{Cam19,AW16,KM16, Scu11,Bra17,Rah12,RA15,KE18},  entanglement\cite{Zhang07,Wangwu19, Huang09, Fun13}, and  correlations\cite{JO02,LJ11,JJ13}, and quantum measurements\cite{ Hor14,Bra15, Kam16, Cha21, Su21}.   Quantum coherence, for instance, was observed in recent  
, experimental realizations  of quantum heat engines based on  nuclear magnetic resonance \cite{RJ19, Ser19}  and  nitrogen-vacancy centers in diamond\cite{JK19}. For quantifying the coherence, several measurements were recently discussed\cite{Fra19,TM14,AG17,XY15,IM16}; among them, the so-called relative entropy of coherence was usually used in quantum thermal machines as it can give quantitative relations between coherence generation and  irreversibility due to finite-time operation of engines\cite{TM14, Fra19}.  For a finite-time unitary driving, the irreversible work comes from not only the coherence but also the incoherent transitions among the energy levels \cite{Fra19}.  A growing interest has been attracted in understanding the thermodynamic implication of the cyclic quantum thermal machine in which unitary driving strokes are involved \cite{Kos02, Pla14, Ale15, Tur19, Jiao21, Cam19,Abah12}. 
 These studies showed the finite-time performance of the intrinsically irreversible quantum heat engines, and also revealed the deep connection  between coherence and quantum friction  responsible for the diabatic transitions \cite{Cam19}.  
 
 Apart from the standard cyclic engines working with two thermal baths, heat engines may be fulled by miniaturized, nonthermal baths which drive systems far away from thermal equilibrium.   The nonthermal baths may be quantum coherent\cite{HT06,Scu11},
quantum correlated\cite{RD09,Ber17,And19,Per15}, quantum-measurement-induced\cite{LB19,CE17,CE18, Su21}, and squeezed\cite{Aba14, Wang19, Nie16, Manz18, Manz16, You18, Sin20,Kla17, Liu15}. 
 Quite naturally, recent studies have started studying  novel finite-time performance and the fluctuations in quantum heat engines subjected to nonthermal baths. Adopting  squeezed baths are particularly interesting  for quantum Otto engines \cite{Kla17, Aba14, Wang19,Assis20} where novel performance beyond the conventional engines with no violation of thermodynamic principles.  

For microscopic systems, heat and work are no longer deterministic \cite{Sek10, Sei12, Smi18} as is the case for macroscopic systems. As a result, the efficiency and power for quantum heat engines are stochastic, and both of them are fluctuating.  The power
fluctuations, together with the efficiency fluctuations, as a
limiting factor for the practical usefulness in heat engines, measure the machine stability \cite{ Bou21}. Ideally, the quantum heat engine should have  high efficiency (small entropy production), large power, and  small fluctuations for these performance measures. Nevertheless, the machine performance and fluctuations are always governed by the thermodynamic uncertainty relation associated with tradeoff between relative fluctuations and irreversible entropy production\cite{Ma18, Sei18}. In that context,  strong emphasis has been put
on the finite-time performance of the quantum heat engines, and in particular on fluctuations of power and efficiency \cite{Ver14, Jiang15, Lut20,Lut21}. On the other hand,  it has been shown that the efficiency for the standard heat engines can be higher than the Carnot value, the least likely in the long-time limit\cite{Ver14, Lut21}. 

In this paper, we investigate the performance and fluctuations in a finite-time quantum Otto engine fuelled by a squeezed bath. We determine the probability distribution functions for heat and work for the engine cycle by using full counting statistics. We then present general formulae of the thermodynamic efficiency, power and power fluctuations, in which  the irreversiblities are associated  both with quantum friction and coherence generated along the finite-time unitary driving  and with system-bath interaction  involved.  By  a numerical simulation with
a qubit employed in the experiment, we show that   the squeezing suppresses the coherence contributing to the unwanted irreversible work and thus results in enhancing the power and thermodynamic efficiency.    
We specifically study the relative  power fluctuations, which characterize the instability of the quantum engines, and the large deviation function of efficiency. Our results show  that reservoir squeezing can significantly reduce the relative power fluctuations and quicken the  convergence speed of the efficiency to the typical value.

\section{Quantum Otto cycle in finite time}
\subsection{Machine model}
We consider a quantum Otto engine cycle working between a hot squeezed and a cold thermal bath [see Fig. \ref{model}(a)]. This engine cycle consists of two unitary driving strokes, where the system is isolated from two heat reservoirs, and two isochoric branches, along each of which  the system with constant Hamiltonian is  weakly coupled to the hot squeezed or cold thermal   bath. Such an engine model is described as follows.

(1){Adiabatic compression from time $t=t_0$ to $t=t_1$. For simplicity, we assume that the initial time $t_0=0$.    During this stroke where the system is isolated from a heat reservoir in time duration $\tau_{ch}$ with $\tau_{ch}=t_1$, the system energy gap is  enlarged by the driven
Hamiltonian $H_{ch}(t)=\frac{\hbar\omega(t)}{2}[\cos(\frac{\pi t}{2\tau_{ch}})\sigma_{x}+\sin(\frac{\pi t}{2\tau_{ch}})\sigma_{z}]$,
where $\omega(t)=\omega_{c}(1-{t}/\tau_{ch})+\omega_{h}({t}/{\tau_{ch}})$ with $0\le t\le\tau_{ch}$, and  $\sigma_{x,y,z}$ are the Pauli matrices.  As this driven Hamiltonian 
does not commute at different times, the quantum coherence is generated in the energy basis of the system. We use this protocol design because it was employed in recent experimental realization of the quantum Otto engines\cite{Ser19,RJ19}. 
The system dynamics can be described by a unitary evolution, provided that the driven time is small enough in order for the energy exchange between the system and its environment to be neglected. Hence, the state of the system $\rho_t$, with $\rho_t\equiv\rho(t)$,  is described in terms of the initial state $\rho_{t_0}$,
\begin{equation}\label{rhob}
\rho_{t}=U_{ch}\rho_{t_0}U_{ch}^{\dag}, 
\end{equation}
where the time evolution operator $U_{ch}= \mathcal{T}_{>}\exp\{-\frac{i}{\hbar}\int_{t_0}^{t_1}dt H_{ch}(t)\}$, with the time-ordering operator $\mathcal{T}_{>}$. In a realistic scenario, we consider that the interaction between the system and the external filed is weak and  the change of internal energy is equal to the work extracted from the system. We then use  full counting statistics method \cite{Sol15, Xu18} to obtain the work distribution as
\begin{eqnarray}\label{pwch}
 p(\mathrm{w}_{ch})&=&\sum_{n,n^{\prime},m}U_{ch}^{mn}\langle n(t_0)|\rho_{t_0}|n^{\prime}(t_0)\rangle U_{ch}^{n'm\dag}\nonumber\\
 &\times&\delta\left[\mathrm{w}_{ch}-\left(\varepsilon_{m}^{h}-\frac{\varepsilon_{n}^{c}+\varepsilon_{n^{\prime}}^{c}}{2}\right)\right]
  \end{eqnarray}
where we have used $H_{ch}(t_0)|n(t_0)\rangle=\varepsilon_n^c |n(t_0)\rangle$, $H_{ch}(t_0)|n^{\prime}(t_0)\rangle=\varepsilon_{n^{\prime}}^c |n^{\prime}(t_0)\rangle$, $H_{ch}(t_1)|m(t_1)\rangle=\varepsilon_m^h |m(t_1)\rangle$, $U_{ch}^{mn}=\langle m(t_1)|U_{ch}|n(t_0)\rangle$, and $U_{ch}^{n^{\prime}m\dag}=\langle n^{\prime}(t_0)|U_{ch}^{\dag}|m(t_1)\rangle$. Here the term $|\langle n(t_0)|U_{ch}|m(t_1)\rangle|^{2}$  denotes the transition probability between the system eigenstates $|n(t_0)\rangle$ and $|m(t_1)\rangle$\cite{RJ19}, and $\langle n(t_0)|\rho_{t_0}|n(t_0)\rangle$ is the probability of the system being in state $|n(t_0)\rangle$. For the two-level system under consideration, we use $\big|g\rangle$ and $\big|e\rangle$ $(n, m=g,e)$ to denote the ground and excited eigenstates, respectively. 
When the driven stroke is a quantum adiabatic, transition between any two eigenstates will not happen  and therefore $|\langle n(t_0)|U_{ch}|m(t_1)\rangle|^{2}=\delta_{nm}$. 
However, along the finite-time driven stroke, the inner friction causes the transitions among the instantaneous energy eigenstates, resulting in  the  irreversible work.

\begin{figure}
\includegraphics[width=3in]{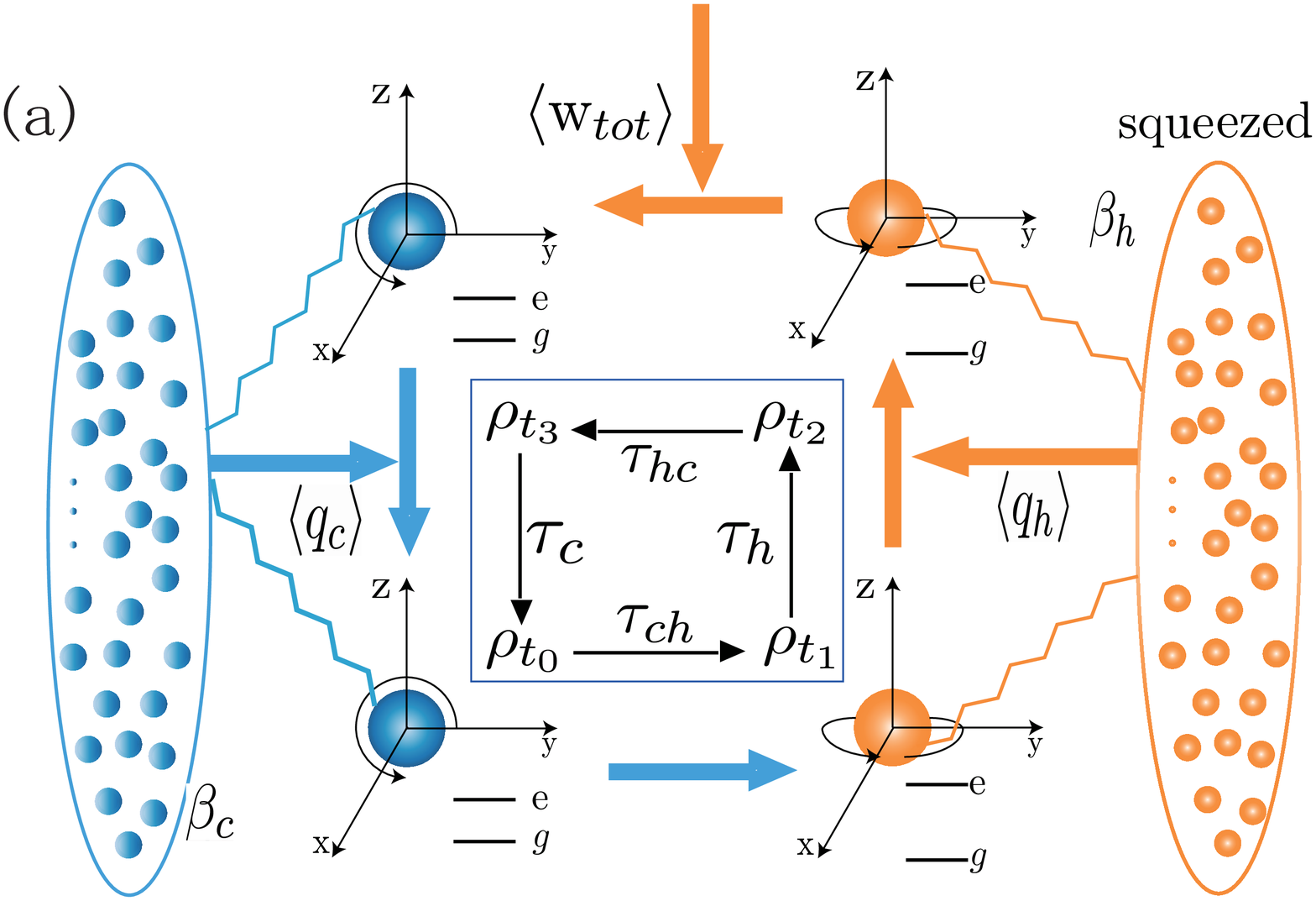} 
\includegraphics[width=3in]{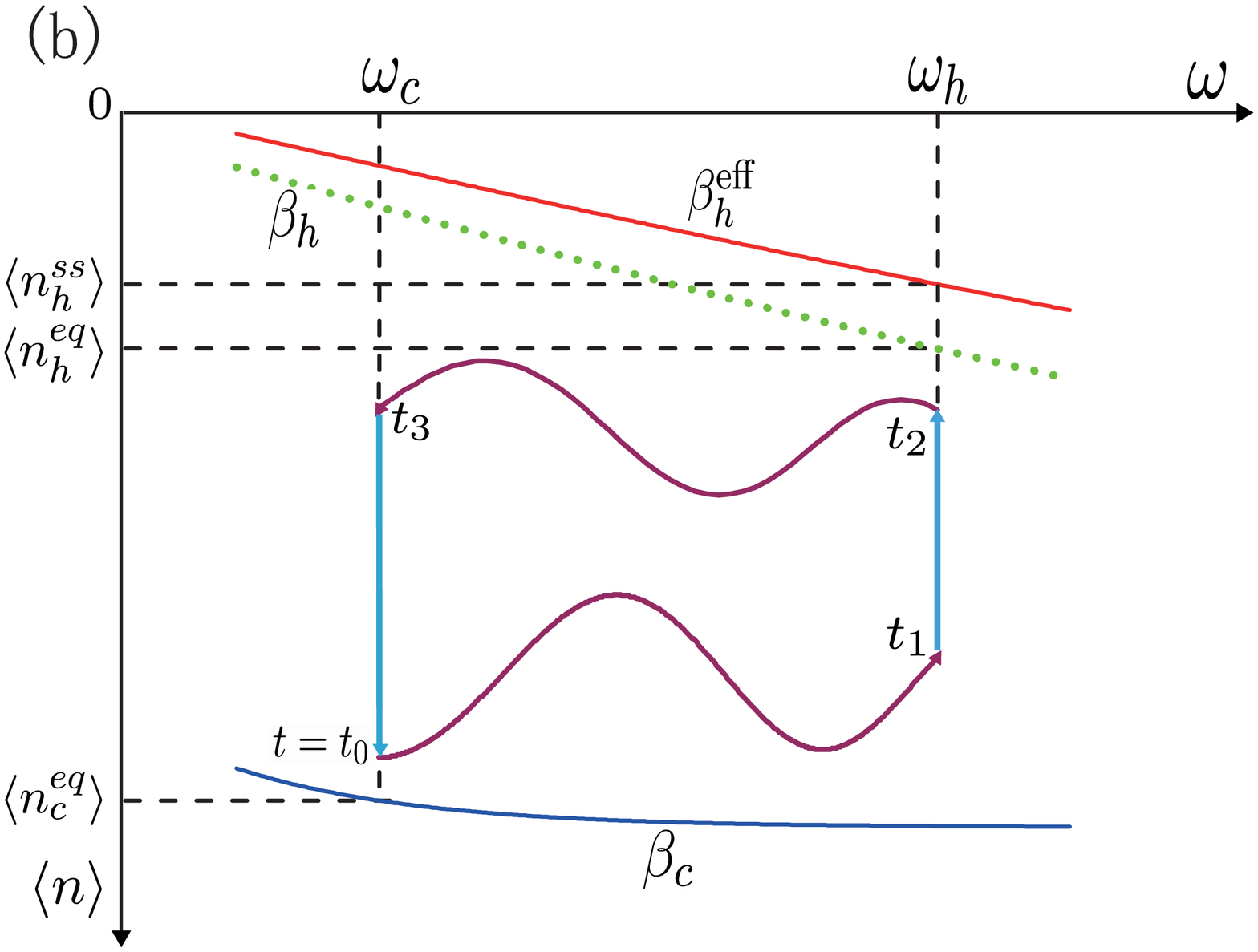}
\caption{ (a) Illustration of the engine cycle driven by a squeezed reservoir. Beginning with  a state $\rho_{t_0}$ at time $t=t_0$, the working substance undergoes a unitary compression mediated by a time-dependent Hamiltonian in  time duration $\tau_{ch}$. In the second stroke from time $t=t_1$ to $t=t_2$, the working substance with its constant Hamiltonian is coupled to a squeezed bath with inverse temperature $\beta_h$ in  time $\tau_h$.   The third stroke is  a unitary expansion where the system state   evolves from $\rho_{t_2}$ to $\rho_{t_3}$. 
  In the fourth stroke, where the system Hamiltonian is again kept constant, the working substance in contact with the cold thermal bath of inverse temperature $\beta_c$  relaxes to
the initial  state after time duration $\tau_{c}$. (b) Schematic diagram of the engine cycle in the ($\omega, \langle n \rangle$) plane. Here $\langle n_h^{{ss}}\rangle=[\exp(\beta_h^{\mathrm{eff}})-1)]$ is excitation number of the system reaching the steady state under reservoir squeezing, and $\langle n_c^{eq}\rangle$ ($\langle n_h^{eq} \rangle$) is excitation
number of the system at thermal equilibrium with the cold (hot) thermal bath of inverse temperature $\beta_h(\beta_c)$. We see that $\langle n_{t_2}\rangle (\langle n_{t_0}\rangle)$ deviates from its asymptotic value $\langle n_h^{ss}\rangle$ ($\langle n_c^{eq}\rangle$) due to finite time operation.}
\label{model}
\end{figure}
(2) {Isochoric heating} from time $t=t_1$ to $t=t_2$. The two-level system is weakly coupled to a hot squeezed heat reservoir
at inverse temperature $\beta_{h}$ during  time duration $\tau_h$ with $\tau_h=t_2-t_1$, while its Hamiltonian  keeps constant:
$H_h(t)=H_{ch}(t_1)=\frac{\hbar \omega_{h}}{2}\sigma_{z}$. As no work is produced along the isochoric process, the  stochastic heat injection is equivalent to the increase of the system eigenenergy.  The transition probability from eigenstate $k$ to $l$ along the hot isochore, $\big|\langle k(t_1) |U_h(t)|l(t_2)\rangle\big|^2$, with the time evolution operator $U_{h}(t)$, can be obtained by using    $\rho_t=U_h \big|k(t_1)\rangle\langle k(t_1)\big| U_h^\dag$ to arrive at  
$\big|\langle k(t_1) |U_h(t)|l(t_1)\rangle\big|^2=\langle l\big|\rho_t|l\rangle$.
Using $H_h(t)\big|l(t_2)=\varepsilon_l^h|l(t_2)\rangle$ and  $H_h(t)\big|k(t_1)=\varepsilon_k^h|k(t_1)\rangle$, it follows that,
the  probability distribution for the heat absorbed during this stroke  reads 
\begin{equation}\label{QH}
p(q_{h}|\mathrm{w}_{ch})=\sum_{k,l}\delta[q_{h}-(\varepsilon_{l}^{h}-\varepsilon_{k}^{h})]\langle l(t_2) \big| \rho_{t_2}\big|l(t_2)\rangle\delta_{k m},
\end{equation}
where  $\langle l(t_2)\big| \rho_{t_2}\big|l(t_2)\rangle$ is the  probability of finding  the system to be in eigenstate $|l(t_2)\rangle$ after the second projective measurement at the isochore. 

In the hot isochoric process within   $t_{1} \le t\le t_2$,  where  the system weakly  interacts with a squeezed thermal bath, the  system  dynamics of density operator $\rho_t$ in
interaction picture   \cite{HF02,MM97,Sri18} can be described by
\begin{equation}\label{drho}
\begin{aligned}
  \frac{d}{dt}\rho_t&=\gamma_{h}(N_h^{ss}+1)(\sigma_{-}\rho_t\sigma_{+}-\frac{1}{2}\sigma_{+}\sigma_{-}\rho_t\\ 
  &-\frac{1}{2}\rho_t\sigma_{+}\sigma_{-})
  +\gamma_{h}N_h^{ss}(\sigma_{+}\rho_t\sigma_{-}\\ 
  &-\frac{1}{2}\sigma_{-}\sigma_{+}\rho_t
  -\frac{1}{2}\rho_t\sigma_{-}\sigma_{+})\\ 
  &-\gamma_{h}M\sigma_{+}\rho_t\sigma_{+}-\gamma_{h}M^{*}\sigma_{-}\rho_t\sigma_{-},
  \end{aligned}
\end{equation}
 where
 \begin{eqnarray}\label{N}
   N_h^{ss}=N_{h}^{th}[\cosh^{2}(r)+\sinh^{2}(r)]+\sinh^{2}(r)
 \end{eqnarray}
 denotes the excitation number of the system which reaches the steady state with the squeezed bath, with the mean population at thermal state  $N_{h}^{th}=\frac{1}{e^{\beta_{h}\hbar w_{h}}-1}$,   $M=-\frac{1}{2}\sinh(2*r)e^{i\theta}(2N_{h}^{th}+1)$, with the phase factor $\theta$, and  $\gamma_h$ is the  vacuum decay rate which indicates the system-bath interaction strength. Here and hereafter we use $\sigma_{+}=|1\rangle\langle0|=\frac{1}{2}(\sigma_{x}+i\sigma_{y})$ and  $\sigma_{-}=|0\rangle\langle1|=\frac{1}{2}(\sigma_{x}-i\sigma_{y})$.

From Eq. (\ref{drho}), one can find \cite{Sri18} that the time-dependent state  $\rho(t)$ along the hot isochore can be determined in the Schr\"{o}dinger picture according to
\begin{eqnarray}\label{rhot2}
\rho_t=
\begin{pmatrix}
\frac{1+\langle\sigma_{z}(t)\rangle}{2}&\mathcal{X}e^{-i\omega_{h}t}\\
\mathcal{X}^{*}e^{i\omega_{h}t}&\frac{1-\langle\sigma_{z}(t)\rangle}{2}
\end{pmatrix},
\end{eqnarray}
where
 $\mathcal{X}=(1+\frac{e^{\gamma_{h}at}-1}{2})e^{-\frac{\gamma_{h}(2N^{ss}_{h}+1+a)t}{2}}\langle\sigma_{-}(t_1)\rangle+\sinh(\frac{\gamma_{h}at}{2})e^{i\Phi-\frac{\gamma_{0}(2N^{ss}_h+1)t}{2}}\langle\sigma_{+}(t_1)\rangle,$  with $a=\sinh(2r)(1+2N_{h}^{th})$, and $ \langle\sigma_{z}(t)\rangle=e^{-\gamma_{h}(2N^{ss}_h+1)t}\langle\sigma_{z}(t_1)\rangle-\frac{1-e^{-\gamma_{h}(2N^{ss}_h+1)t}}{2N^{ss}_h+1}$.

When  the isochoric stroke is slow enough such that $\tau_{h}\gg\tau_{h,relax}$, where $\tau_{h,relax}$
is the relaxation time
of the system with the hot bath, the density matrix (\ref{rhot2})  reduces to 
\begin{equation}\label{rhoc}
  \rho_{t_2}^{\mathrm{ss}}\Big|_{\tau_h\gg\tau_{h,relax}}=
  \begin{pmatrix}
   p_{t_2}^{e,ss} &0\\
   0&p_{t_2}^{g,ss}
   \end{pmatrix}
\end{equation}
where $p_{t_2}^{e,ss}=N_{h}^{ss}/(2N_h^{ss}+1)$ and $p_{t_2}^{g,ss}=1-p_{t_2}^{e,ss}$.
 If we introduce the effective inverse temperature 
 \begin{equation}
  \beta_h^{\mathrm{eff}}=\frac{1}{\hbar\omega_h}\ln\frac{N_{h}^{ss}+1}{N_{h}^{ss}}   \label{betasq}
 \end{equation}
to write the excitation number as $N_h^{ss}=1/(e^{\beta_h^{\mathrm{eff}}\omega_h}-1)$,  the detailed balance is restored in the squeezing case owing to the relation  $\frac{\langle e(t_2)\big|\rho_{t_2}\big|e(\tau_{t_2})\rangle}{\langle g(t_2)\big|\rho_{t_2}\big|g(t_2)\rangle}\Big|_{\tau_h\gg\tau_{h,relax}}=e^{-\beta_h^{\mathrm{eff}}\omega_h}$. When  $\tau_h\gg\tau_{h,relax}$,   the system could be fully thermalized along the hot isochore and all the coherence produced in the compression would be erased [see Eq. (\ref {rhoc})]. However,  for incomplete thermalization with $\tau_h\le\tau_{h,relax},$  a residual amount of the coherence  is retained  and thus the parameter $\mathcal{X}$ in Eq. (\ref{rhot2}) is nonzero. Such coherence that endures in this finite-time isochoric process will be present in the next driven stroke.

(3) {Adiabatic expansion} from time $t=t_2$ to $t=t_3$. The stroke, during which  the driven Hamiltonian $H_{hc}(t_3-t)=H_{ch}(t)$  is realized by reversing the protocol used in the adiabatic compression, such that the expansion Hamiltonian takes the values the same as the compression  Hamiltonian, namely, $\tau_{hc}=\tau_{ch}$. 
Since the evolution is unitary, we obtain
\begin{equation}\label{PD}
  \rho_{t_3}=U_{hc}\rho_{t_2}U_{hc}^{\dag},
\end{equation}
where $U_{hc}=\mathcal{T}_{>}\exp\{-\frac{i}{\hbar}\int_{t_2}^{t_3}dt H_{hc}(t)\}$.
Like the first stroke, the internal friction (due to finite time evolution) will bring coherence and  results in the transition between the  eigenstates of the Hamiltonian
$H_{hc}(t_3)$ and $ H_{hc}(t_2) $. The quantum work distribution along the driven stroke can be expressed  as 
\begin{equation}\label{pwhc}
\begin{aligned}
  p(\mathrm{w}_{hc}|\mathrm{w}_{ch},q_{h})&=\sum_{i,i^{\prime},j}U_{hc}^{j,i}\langle i(t_2)|\rho_{t_2}|i^{\prime}(t_2)\rangle U_{hc}^{i^{\prime}j\dag}\\
 & \times\delta[\mathrm{w}_{hc}-(\varepsilon_{j}^{c}-\frac{\varepsilon_{i}^{h}+\varepsilon_{i\prime}^{h}}{2})]\delta_{i,l}, 
  \end{aligned}
\end{equation}
where we have used $H_{hc}(t_2)|i(t_2)\rangle=\varepsilon_i^h |i(t_2)\rangle$, $H_{hc}(t_2)|i^{\prime}(t_2)\rangle=\varepsilon_{i^{\prime}}^h |i^{\prime}(t_2)\rangle$, $H_{hc}(t_3)|j(t_3)\rangle=\varepsilon_j^c |j(t_3)\rangle$, $U_{hc}^{ji}=\langle j(t_3)|U_{hc}|i(t_2)\rangle$, and $U_{hc}^{i^{\prime}j}=\langle i^{\prime}(t_2)|U_{hc}^{\dag}|j(t_3)\rangle$. The term $|\langle i(t_2)|U_{hc}|j(t_3)\rangle|^{2}$  is
the transition probability from state $|i(t_2)\rangle$  to $|j(t_3)\rangle$, and $\langle i(t_2)|\rho_{t_2}|i'(t_2)\rangle\delta_{i,i'}$ denotes the probability for finding  the system  in state $|i\rangle$.

(4) {Isochoric cooling} from $t=t_3$ to $t=t_3+\tau_c$.  The system is weakly coupled with a cold thermal at inverse temperature $\beta_{c}$ in time period $\tau_c$, and its  Hamiltonian is kept constant at $H_{c}(t)$=$H_{ch}(0)=\frac{\omega_{0}}{2}\sigma_{x}$.  The dynamics of the state $\rho_t$ along this stroke, where no squeezing is present and $r=0$, can be given in interaction picture by 
\begin{equation}\label{drhoc}
\begin{aligned}
  \frac{d}{dt}\rho_t&=\gamma_{c}(N_{c}^{th}+1)[\sigma_{-}\rho_t\sigma_{+}-\frac{1}{2}\sigma_{+}\sigma_{-}\rho_t\\ 
  &-\frac{1}{2}\rho_t\sigma_{+}\sigma_{-}]
  +\gamma_{c}N_{c}^{th}[\sigma_{+}\rho_t\sigma_{-}\\ 
  &-\frac{1}{2}\sigma_{-}\sigma_{+}\rho_t
  -\frac{1}{2}\rho_t\sigma_{-}\sigma_{+}],
  \end{aligned}
\end{equation}
 where $N_{c}^{th}=\frac{1}{e^{\beta_{c}\hbar w_{c}}-1}$ is the excitation number of the system at thermal equilibrium with the cold bath, and $\gamma_c$ is the interaction strength between the system and the cold thermal bath.
Using Eq. (\ref{drhoc}) , we can obtain the density operator $\rho_t $ of the system along the cold isochore, and  then make a unitary transformation of the
density matrix  from the $\sigma_z$ basis to the $\sigma_x$ basis to arrive at 
\begin{equation}\label{rhot3}
\rho_t=\mathcal{U}
\begin{pmatrix}
\frac{1+\langle\sigma_{z}(t)\rangle}{2}&\mathcal{Y}e^{-i\omega_{c}t}\\
\mathcal{Y}^{*}e^{i\omega_{c}t}&\frac{1-\langle\sigma_{z}(t)\rangle}{2}
\end{pmatrix}\mathcal{U}^{\dag},
\end{equation}
where the transformation matrix $\mathcal{U}=\begin{pmatrix}
\langle e(t_1)|e(t_0)\rangle&\langle g(t_1)|e(t_0)\rangle\\
\langle e(t_1)|g(t_0)\rangle&\langle g(t_1)|g(t_0)\rangle
\end{pmatrix}$ and 
 $\mathcal{Y}=e^{-\frac{\gamma_c(2N_{c}^{th}+1)t}{2}}\langle\sigma_{-}(t_3)\rangle$.
 We assume that the final density matrix of the stroke is equal to
$\rho_{t_0}$ for closing the engine cycle. When the time duration $\tau_c$ is much larger than the relaxation time $\tau_{c,relax}$, $\mathcal{Y}$ becomes vanishing and the system reaches the thermal equilibrium  at the end of the cooling stroke, and then the state of the system takes the form (in the $\sigma_x$ basis):
\begin{equation}\label{rhoth}
  \rho_{t_0}\Big|_{\tau_c\gg\tau_{c,relax}}=
  \begin{pmatrix}
   \frac{1}{2}&\frac{p_{t_0}^{e,eq}-p_{t_0}^{g,eq}}{2}\\
   \frac{p_{t_0}^{e,eq}-p_{t_0}^{g,eq}}{2}&\frac{1}{2}
   \end{pmatrix},
\end{equation}
where $p_{t_0}^{e,eq}=N_c^{eq}/(2N_c^{eq}+1)$ and $p_{t_0}^{g,eq}=1-p_{t_0}^{e,eq}$. 

\subsection{Statistics of machine performance parameters}
 The stochastic   work extracted from
 the system in a single cycle is -$\mathrm{w}_{tot}=-(\mathrm{w}_{ch}+\mathrm{w}_{hc})$, and the stochastic efficiency reads $\eta=-\mathrm{w}_{tot}/q_{h}$. The joint distribution \cite{Hol18} for the  work output $\mathrm{w}_{ch}$, $\mathrm{w}_{hc}$ and heat $q_{h}$ can be determined according to Eqs. (\ref{pwch}), (\ref{QH}), and (\ref{pwhc}) 
\begin{eqnarray}\label{wqhw}
  p(\mathrm{w}_{ch},q_{h},\mathrm{w}_{hc})&=&p(\mathrm{w}_{2}|\mathrm{w}_{ch},q_{h})p(q_{h}|\mathrm{w}_{ch})p(\mathrm{w}_{ch})\nonumber\\
  &=&\sum_{n,n^{\prime} m,i,i^{\prime},j}\delta[\mathrm{w}_{ch}-(\varepsilon_{m}^{h}-\frac{\varepsilon_{n}^{c}+\varepsilon_{n^{\prime}}^{c}}{2})]\nonumber\\
  &\times&\delta[q_{h}-(\varepsilon^{h}_{i}-\varepsilon^{h}_{m})]\nonumber\\
  &\times&\delta[\mathrm{w}_{hc}-(\varepsilon_{j}^{c}-\frac{\varepsilon_{i}^{h}+\varepsilon_{i\prime}^{h}}{2})]\nonumber\\
  &\times&U_{ch}^{mn}\langle n(t_0)|\rho_{t_0}|n^{\prime}(t_0)\rangle U_{ch}^{n^{\prime}m}\nonumber\\
  &\times& U_{hc}^{ji}\langle i(t_2)|\rho_{t_2}|i^{\prime}(t_2)\rangle U_{hc}^{i^{\prime}j}.
\end{eqnarray}
This sets the joint distribution of the total work $\mathrm{w}_{tot}$,  
\begin{equation}\label{JD}
   p(q_{h},\mathrm{w}_{tot})=\int d\mathrm{w}_{ch}d\mathrm{w}_{hc}\delta[\mathrm{w}
   -(\mathrm{w}_{ch}+\mathrm{w}_{hc})]p(\mathrm{w}_{ch},q_{h},\mathrm{w}_{hc}),
\end{equation}
for a cycle with quantum heat injection $q_h$.  Let $\xi\equiv |\langle n(t_0)|U_{ch}|m(t_1)\rangle|^{2}=|\langle i(t_2)|U_{hc}|j(t_3)\rangle|^{2} $ ( $m,n,i,j=e,g$)  be 
 the level transition probability during expansion or compression for the two level system.  Using Eq. (\ref{JD}),  we obtain (see Appendix \ref{appower} for details)  average work $\langle \mathrm{w}_{tot}\rangle $ and average injection $\langle q_h\rangle$ as 
\begin{eqnarray}\label{wtot}
-\langle\mathrm{w}_{tot}\rangle&=&\hbar(w_{h}-w_{c})(\langle n_{t_2}\rangle-\langle n_{t_0}\rangle)\nonumber\\
&+&2\hbar\xi(w_{c}\langle n_{t_2}\rangle+w_{h}\langle n_{t_0}\rangle)\nonumber\\
&-&
2\hbar w_{h}\zeta_{ch}-2\hbar w_{c}\zeta_{hc}, \label{wtot}
\end{eqnarray}
\begin{eqnarray}\label{qh}
\langle q_{h}\rangle=\hbar w_h[\langle n_{t_2}\rangle+\langle n_{t_0}\rangle(2\xi-1)-2\zeta_{ch}],
\end{eqnarray}
where we have used
$2\langle n_{t_0}\rangle:=-\langle g(t_0)|\rho_{t_0}|g(t_0)\rangle+\langle e(t_0)|\rho_{t_0}|e(t_0)\rangle$, $2\langle n_{t_2}\rangle:=-\langle g(t_2)|\rho_{t_2}|g(t_2)\rangle+\langle e(t_2)|\rho_{t_2}|e(t_2)\rangle$, $\zeta_{ch}:=-Re[U_{ch}^{gg}\langle g(t_0)|\rho_{t_0}|e(t_0)\rangle U_{ch}^{eg\dag}]$, and  $\zeta_{hc}:=-Re[U_{hc}^{gg}\langle g(t_2)|\rho_{t_2}|e(t_2)\rangle U_{hc}^{eg\dag}]$.  After a single cycle, the working substance  produces the total work $-\langle \mathrm{w}_{tot}\rangle$ by absorbing average heats from the hot and cold baths, $\langle q_h\rangle$ and $\langle q_c\rangle$, where $\langle q_c\rangle=-\langle \mathrm{w}_{tot}\rangle-\langle q_h\rangle$ due to the energy conservation, as sketched in Fig. \ref{model}(b). Here $\langle n_{t_0}\rangle$ and  $\langle n_{t_2}\rangle$ are the average populations of  the two-level system at times $t=t_0$ and $t=t_2$ [see also Fig. \ref{model} (b)], respectively.  

The parameters $\zeta_{hc}$ and $\zeta_{ch}$ in Eq. (\ref{wtot}), associated with off-diagonal elements,  refer to the quantum coherence between thee two energy  eigenstates. 
These two parameters  are not independent but are correlated  for finite-time, incomplete thermalizations where coherence is not fully erased.  $\xi_{ch}$ (determined by the off-diagonal elements)    affects the state $\rho_{t_1}$ which evaluates to $\rho_{t_2}$ after the hot thermal contact in finite time, and thus $\xi_{hc}$ is  dependent on $\xi_{ch}$; similarly, $\xi_{ch}$ is affected by $\xi_{hc}$ due to finite time $\tau_c$. It is therefor indicated  that $\xi_{ch}$ and  $\rho_{hc}$ are correlated with each other.   
The residual coherence, transferred from the first to the third strokes due to incomplete thermalization, would interfere with the coherence generated in the third stroke. To reveal such  a dynamical interference effect on the thermodynamic quantities of the machine, our quantum heat engine is compared with an alternative cycle, where a full dephasing operation is performed to completely remove all coherence after the hot isochore with any value of thermal-contact time $\tau_h$. 
Throughout the paper we use the superscript
``deph'' to describe the quantities corresponding to the dephased engine cycle.

The average work -$\langle \mathrm{w}_{tot}\rangle$ (\ref{wtot})
 can be split up into the two terms: 
 \begin{equation}
-\langle 
\mathrm{w}_{tot}\rangle=\langle \mathrm{w}_{\mathrm{deph}}\rangle+\langle \mathrm{w}_{\mathrm{coh}}\rangle, \label{wtoh}
\end{equation}
where 
\begin{equation}
    \langle \mathrm{w}_{\mathrm{deph}}\rangle
    =\hbar(w_{h}-w_{c})(\langle n_{t_2}\rangle-\langle n_{t_0}\rangle)+
    \langle \mathrm{w}_{\mathrm{fri}}\rangle \label{wdep}
\end{equation} with $\langle \mathrm{w}_{\mathrm{fri}}\rangle=:2\hbar\xi(w_{c}\langle n_{t_2}\rangle+w_{h}\langle n_{t_0}\rangle)$, is the average work in the dephased  case, and
\begin{equation}
 \langle{w}_{\mathrm{coh}}\rangle=-2\hbar w_{h}\zeta_{ch}-2\hbar w_{c}\zeta_{hc}  \label{wcoh}
\end{equation}  is the average work associated with quantum coherence. The second term in Eq. (\ref{wdep}), $\langle w_{\mathrm{fri}}\rangle $, represents the additional work due to overcome the inner friction causing unwanted diabatic transitions in instantaneous  energy eigenstates. 
The irreversible work can be decomposed into two contributions, which come from  the inner friction associated with the diabatic transitions  and  the coherence generated  along the unitary driven strokes. 
Due to finite-time operation,  the system at end of either the hot (cold) isochore can not reach the stationary  state (Gibbs thermal state), meaning that $\langle n_{t_2}\rangle <\langle n_h^{ss}\rangle$ and $\langle n_{t_0}\rangle<\langle n_c^{eq}\rangle$  [cf. Fig. \ref{model}(b)]. The first term in Eq. (\ref{wdep}) should monotonically increase with increase of $\tau_h$ and $\tau_c$ to reach a maximum value when $\tau_h\gg\tau_{h,relax}$ and $\tau_c\gg\tau_{c,relax}$ and $\langle n_{t_2}\rangle \rightarrow \langle n_h^{ss}\rangle$ and $\langle n_{t_0}\rangle \rightarrow \langle n_c^{eq}\rangle$. We therefore conclude that the total work for the cyclic heat engine is affected by  the irreversibility  originating from both two  driven and two isochoric processes.

With consideration of Eqs. (\ref{wtot}) and (\ref{qh}), the thermodynamic efficiency, $\eta_{th}=-\langle \mathrm{w}_{tot}\rangle/\langle{q_h}\rangle$, is then obtained as
\begin{equation}\label{eta}
\begin{aligned}
 \eta_{th}
 &=1+\frac{w_{c}}{w_{h}}\frac{\langle n_{t_0}\rangle+\langle n_{t_2}\rangle(2\xi-1)-2\zeta_{hc}}{ \langle n_{t_2}\rangle+\langle n_{t_0}\rangle(2\xi-1)-2\zeta_{ch}}. 
 \end{aligned}
\end{equation}
 Since the times taken   for the thermal contacts and driven strokes are finite, the quantum coherence and quantum friction are 
created,  resulting that the  efficiency (\ref{eta}) 
depends  not only on the inner friction but also on the coherence.
If quantum coherence is negligible ($\zeta_{ch,hc}\rightarrow0$), 
the efficiency reduces to the one\cite{Jiao21},
\begin{equation}
\eta_{th}=1+\frac{w_{c}}{w_{h}}\frac{\langle n_{t_0}\rangle+\langle n_{t_2}\rangle(2\xi-1)}{\langle n_{t_2}\rangle +\langle n_{t_0}\rangle (2\xi-1)}, \label{et11}
\end{equation}
 for the cycle in which the incomplete thermalization  was approximately described by local thermal equilibrium, and it further simplifies to  the one, 
 $\eta_{th}=1-{w_c}/{w_h}$, in the case when inner friction is absent ($\xi\rightarrow0$).  
 
From Eq. (\ref{JD}), we find that the work fluctuations, $\delta \mathrm{w}^{2}=\langle\mathrm{w}^{2}\rangle-\langle \mathrm{w}\rangle^{2}$ can by given by (see Appendix \ref{appower})
\begin{eqnarray}
\delta \mathrm{w}^{2}&=&\hbar^2 w_{h}^{2}\{\frac{1}{2}-\langle n_{t_2}\rangle^2-[\langle n_{t_0}\rangle(1-2\xi)+2\zeta_{ch}]^2\}\nonumber\\
&+&\hbar^2 w_{c}^{2}\{\frac{1}{2}-\langle n_{t_0}\rangle^2-[\langle n_{t_2}\rangle(1-2\xi)+2\zeta_{hc}]^2\}\nonumber\\
&+&\hbar^2 w_{c}w_{h}\{2\langle n_{t_0}\rangle[\langle n_{t_0}\rangle (1-2\xi)+2\zeta_{ch}]\nonumber\\
&+&2\langle n_{t_2}\rangle[\langle n_{t_2}\rangle(1-2\xi)+2\zeta_{hc}]+2\xi-1
\}. \label{wfluc}
\end{eqnarray}
If $\zeta_{ch}=\zeta_{hc}=0$, this simplifies to $\delta\mathrm{w}^2=\hbar^2 w_{h}^{2}[\frac{1}{2}-\langle n_{t_2}\rangle^2-\langle n_{t_0}\rangle^2(1-2\xi)]+\hbar^2 w_{c}^{2}[\frac{1}{2}-\langle n_{t_0}\rangle^2-\langle n_{t_2}\rangle^2(1-2\xi)]+\hbar^2 w_cw_h[2\langle n_{t_0}\rangle^2(1-2\xi)+2\langle n_{t_2}\rangle^2(1-2\xi)+2\xi-1]$, previously derived in Ref. \cite{Jiao21} where no quantum coherence was considered. 

In contrast to the average work, the average efficiency for the quantum Otto engine may be ill-defined due to the possible divergence of the stochastic efficiency. This implies that one could not use the  efficiency variance (or efficiency fluctuations) to investigate the efficiency statistics. We thus resort to large deviation theory \cite{Tou09} associated with the exponential decay of probabilities of large fluctuations, assuming that the quantum engine proceeds in long-time limit. We recall that the large deviation function of the joint distribution $p(q_h, w)$ and the efficiency
distribution $p_{K}(\eta)$ for a large number of cycles ($K\gg1$) are governed by the respective asymptotic forms of $
  p_{K}(q_{h},\mathrm{w}_{tot})\asymp e^{-KI(q_{h},\mathrm{w}_{tot})},
$
 and 
  $p_{K}(\eta)\asymp e^{-KJ(\eta)}.
$ 
The large deviation functions,  $I(q_{h},\mathrm{w}_{tot})$ and $J(\eta)$, describe the exponentially unlikely deviations of  $q_{h}$, $\mathrm{w}_{tot}$ and $\eta$ from their most probable values.
The rate function $J(\eta)$ can be obtained from $I(q_{h},\mathrm{w}_{tot})$ by the contraction: 
$
  J(\eta)=\underset{q_h}{\mathrm{min}}I(q_{h},-\eta q_{h}).
$
Let us define $q_{h}^{(K)}:=\sum^{K}_{j=1}q^{(j)}_{h}/K$ and $\mathrm{w}^{(K)}:=\sum^{K}_{j=1}\mathrm{w}_{tot}^{(j)}/K$
\begin{eqnarray}
 \phi(\varphi_{1},\varphi_{2})&=&\underset{K\rightarrow\infty}{\mathrm{lim}}\frac{1}{K}\ln\langle e^{K(\varphi_{1}q_{h}^{(K)}+\varphi_{2}\mathrm{w}^{(K)})}\rangle\nonumber\\
 &=&\ln\langle e^{\varphi_{1}q_{h}+\varphi_{2}\mathrm{w}_{tot}}\rangle, \label{phi}
\end{eqnarray}
where we have used $\langle e^{\varphi_{1}q_{h}+\varphi_{2}\mathrm{w}_{tot}}\rangle=\int\int dq_{h}d\mathrm{w}_{tot}e^{-\varphi_{1}q_{h}-\varphi_{2}\mathrm{w}_{tot}}p(q_{h},\mathrm{w}_{tot}).
$ It follows, using  the Legendre-Fenchel transform, that the large deviation function of quantum efficiency reads\cite{Lut21, Gver14}
\begin{equation}
  J(\eta)=-\underset{\varphi_{2}}{\mathrm{min}} \phi(\varphi_{2}\eta,\varphi_{2}), \label{jeta}
\end{equation}
where function $\phi$ was defined in Eq. (\ref{phi}).

\section{Explicit connection between efficiency and entropy production }
The Kullback-Leibler-Umegaki (KL) divergence can be used to describe the distance between  an arbitrary state $\rho$ and a reference state $\rho^{\mathrm{ref}}$: $D(\rho||\rho^{\mathrm{ref}})=\mathrm{Tr} (\rho    \mathrm{ln}  \rho)-\mathrm{Tr} (\rho \mathrm{ln}   \rho^{\mathrm{ref}})$\cite{Vedral02,Wilde17}. The KL divergence indicates the degree of nonequilibrium of the state $\rho$ if the reference state $\rho^{\mathrm{ref}}$ is chosen as the thermal state  $\rho^{eq}$. Using   ${\Pi_{n}}:=\sum_n|n\rangle\langle n| \rho|n\rangle\langle n|$ to denote the propagators of $\rho^{\mathrm{ref}}$, the thermal divergence can be rewritten as \cite{TM14} $D(\rho||\rho^{\mathrm{ref}})=D(\mathcal{E}(\rho)||\rho^{\mathrm{ref}})+C(\rho)$, where $\mathcal{E}(\cdot)=\sum_n\Pi_n(\cdot)\Pi_n$ is the dephasing map, removing all coherence in the energy basis. Here 
\begin{equation} C(\rho)=S[\mathcal{E}(\rho)]-S(\rho)~(\mathrm{with}~ S(\rho)=-\mathrm{Tr} [\rho \mathrm{ln}  \rho])
\end{equation} 
is the relative entropy of coherence, which  quantifies the amount of coherence.   

The total entropy production for the quantum Otto engine can be related to the KL divergence along the two thermal contacts through
$\Sigma_{tot}
  =-[\mathrm{Tr}(\rho_{t_2}\ln\rho_h^{ss})-\mathrm{Tr}(\rho_{t_1}\ln\rho_h^{ss}]-[\mathrm{Tr}(\rho_{t_0}\ln\rho_c^{eq})-\mathrm{Tr}(\rho_{t_3}\ln\rho_c^{eq}]$  (for details see Appendix \ref{apeta}), where $\rho_{t_2}$ and $\rho_{t_1}$ ($\rho_{t_0}$ and $\rho_{t_3}$)  are the final and initial instants of the hot (cold) isochore, respectively.  However, this does not mean that the entropy production does not depend on the coherences generated in the unitary expansion and compression whatsoever. The states $\rho_{t_0,t_1, t_2,t_3}$ depend on the coherence generated in the two driven strokes due to finite time driving, and  thus the entropy production $\Sigma_{tot}$ depends on the protocols of  these two processes.   In terms of this entropy production $\Sigma_{tot}$, the machine efficiency can be written as 
\begin{equation}
    \eta_{th}=\eta_{C}^{\mathrm{gen}}-\frac{\Sigma_{tot}}{\beta_{c}\langle q_{h}\rangle}, 
    \label{etage}
\end{equation}
 where $\eta_{C}^{\mathrm{gen}}=1-{\beta_{h}^{\mathrm{eff}}}/{\beta_{c}}$ is the  generalized Cannot efficiency with $\beta_h^{\mathrm{eff}}$ defined in Eq. (\ref{betasq}). In the limit of either high temperature or small squeezing, the effective temperature (\ref{betasq}) becomes $\beta_h^{\mathrm{eff}}=\beta_h\sinh(2 r)$, leading to 
$\eta=\sinh(2r)\eta_C-{\Sigma_{tot}}/({\beta_{c}\langle q_{h}\rangle})$.   
Eq. (\ref{etage})  shows that reservoir squeezing improves the efficiency but the irreversibility captured by the entropy production deteriorates the performance.  
Since the  squeezing   contributes significantly to the efficiency, the upper bound of the efficiency may be beyond the standard Carnot value $\eta_C$ even at positive power where the entropy production is finite. However, the efficiency must be bounded by the generalized Carnot value $\eta_C^{gen}$ due to positive entropy production $\Sigma_{tot}$ and average heat injection $\langle q_h\rangle$, thereby showing that the second law of thermodynamics holds.  The
second term in Eq. (\ref{etage})  quantifies the decrease in the efficiency  due to
the irreversible entropy production which is dependent on the protocol of the finite-time cycle.

\section{Numerical analysis}
 The energy scales of the system will be set to be compatible with experiments based on nuclear magnetic
resonance setups\cite{Ser19}. We choose the energy gaps of the system along the cold and hot isochores  as   $\omega_c= 4\pi\mathrm{kHz}$ and
 $\omega_h =7.2\pi \mathrm{kHz}$, respectively.    The inverse temperatures of the cold and  hot baths are set as $\beta_c=2/(\hbar\omega_c)$ and $\beta_h=1/(2\hbar\omega_h)$. The system-bath interaction strength along  the hot (cold) isochore is assumed as $\gamma_c=\gamma_h=1 \mathrm{Hz}$, from which   the relaxation times are found to be $\tau_{h,relax}=244.92 $ms and $\tau_{c,relax}=761.59 $ms \cite{Cam19}.
\begin{figure}
\includegraphics[width=7cm]{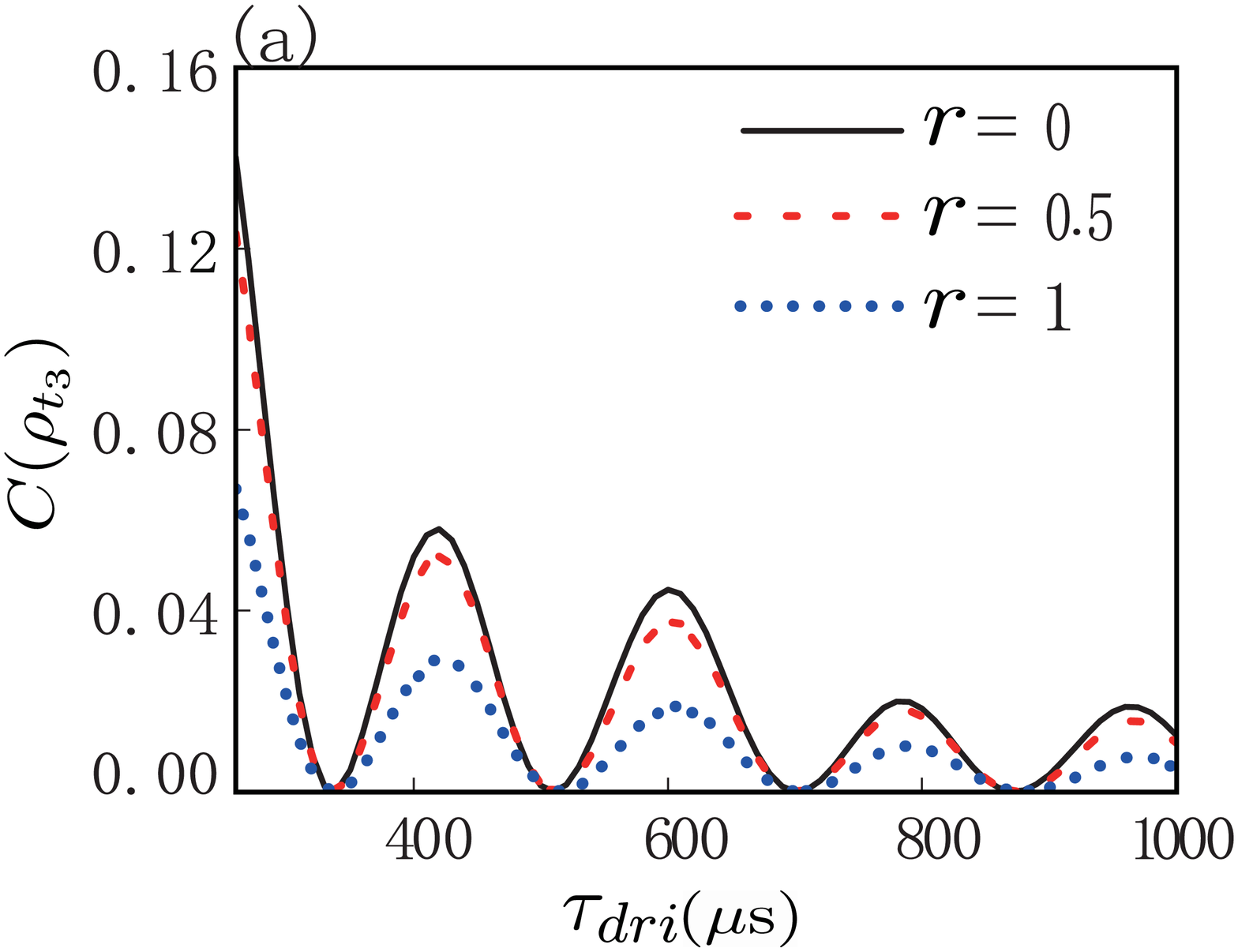}
\includegraphics[width=6.7cm]{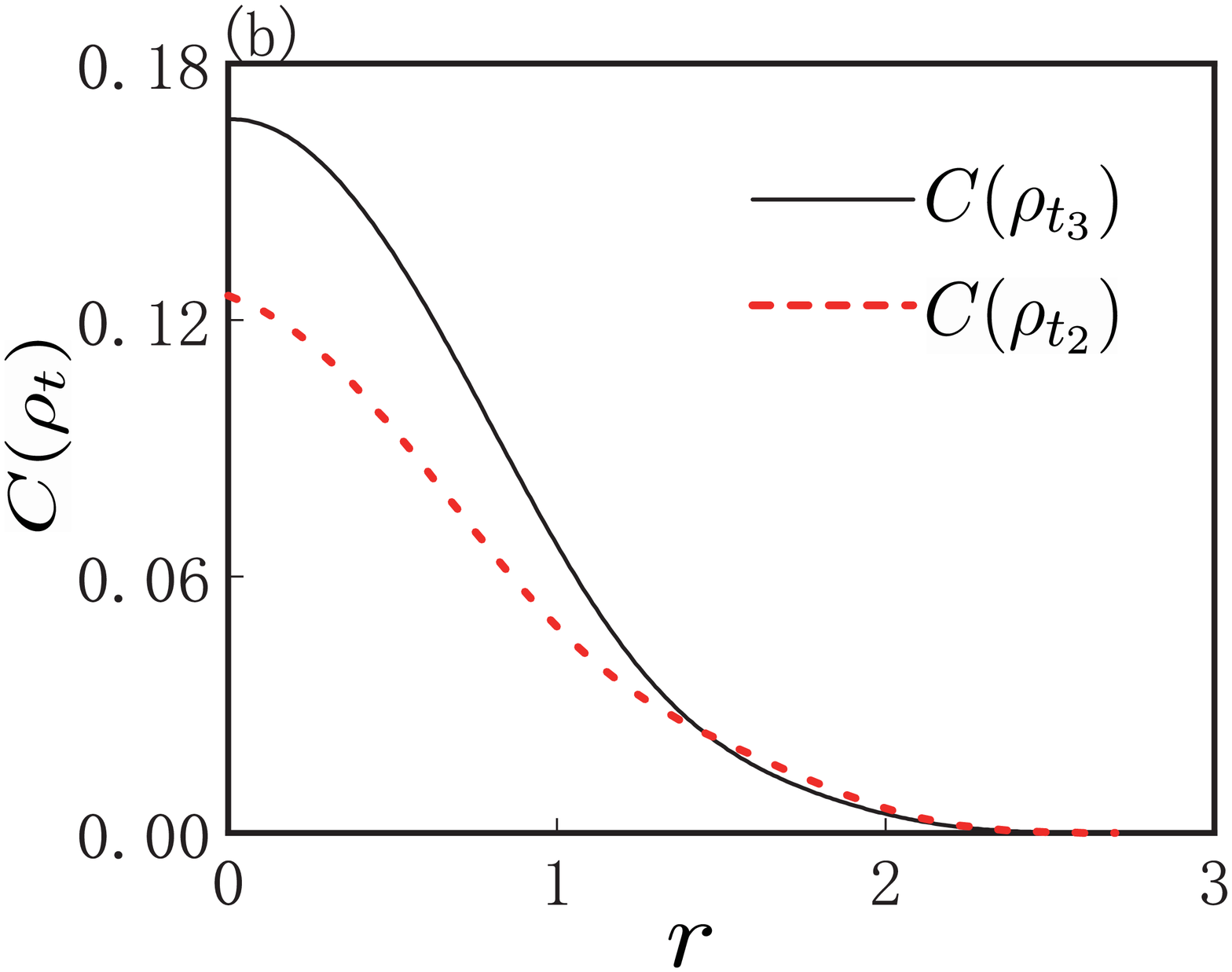}
\caption{  Coherence at state $\rho_{t_3}$ as a function of driving time $\tau_{ch}$ for different values of $r$. (b) Coherence at states $\rho_{t_2}$ and $\rho_{t_3}$  as a function of squeezing parameter $r$  for  $\tau_{ch}=200\mu $s. The time $\tau_h$ is set as $\tau_{h}=0.07515$s.
}
\label{cohd}
\end{figure}
\begin{figure}
{\includegraphics[width=7.48cm]{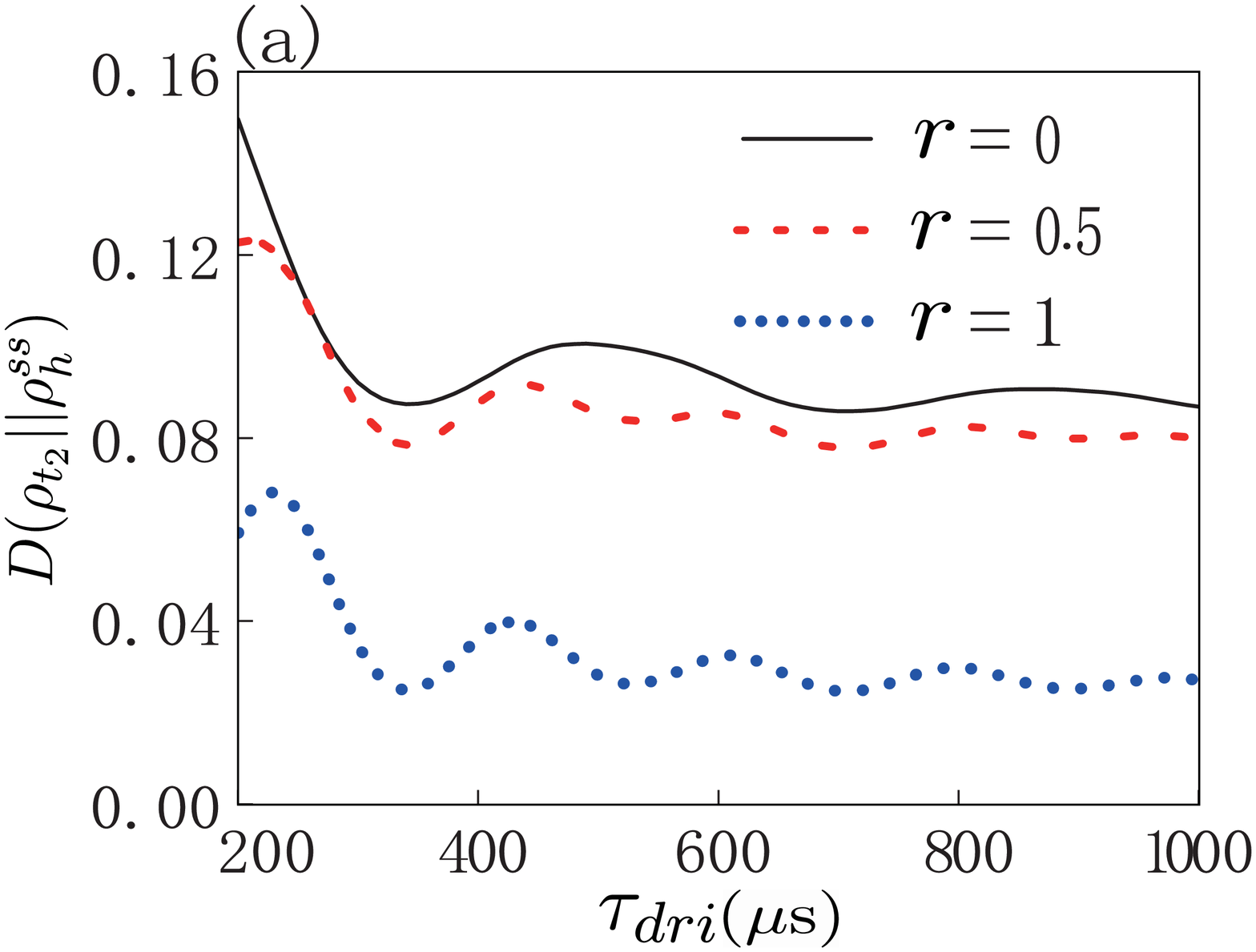}}
{\includegraphics[width=6.98cm]{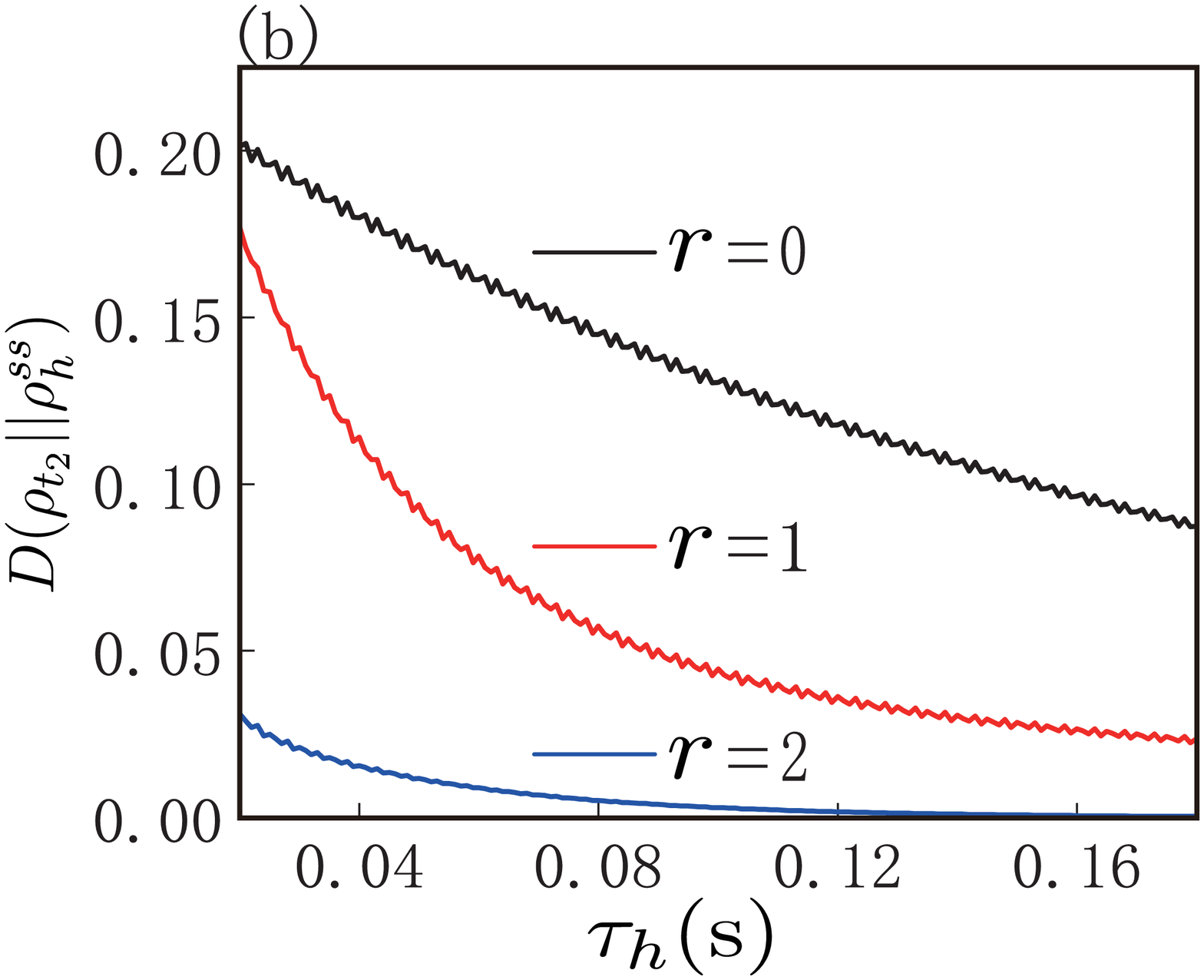}}
\caption{Relative entropy at time $t=t_2$ as a function of  (a) driving time $\tau_{dri}$ and (b) thermal-contact time $\tau_h$ for different values of $r$. In (a)  $\tau_{h}=0.07515$s and in (b)  $\tau_{dri}=200\mu$s.}
\label{lkd}
\end{figure}


 \begin{figure*}
\includegraphics[width=17cm]{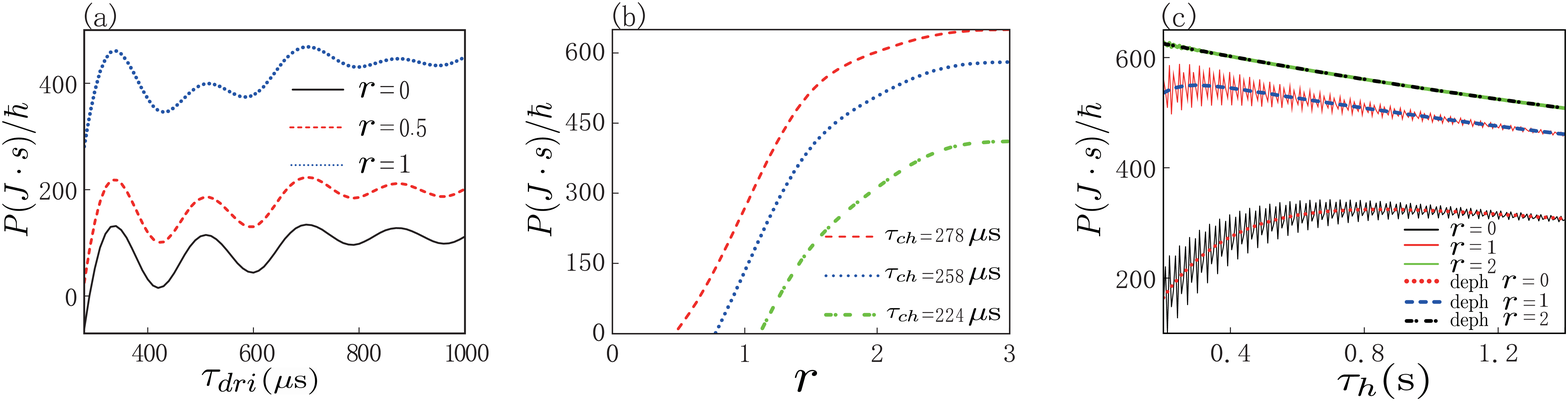}
\caption{Power $P$  in unit of J$\cdot$s$/\hbar$ as a function of  (a) driving time $\tau_{dri}$, (b) squeezing parameter $r$, and (c) thermal-contact time $\tau_h$.   In (a) and (b) $\tau_h=75.15$ms, and in (b) $\tau_{dri}=460\mu$s. In (c) the cases of $r=0,1,2$ in the dephased engine cycle (labelled for``deph'') by red dotted line, blue dashed line, and black  dot-dashed line,  respectively. }
\label{p}
\end{figure*}
The time spent on the hot isochore is smaller than or of the same order as the relaxation time,  $\tau_{h}\lesssim\tau_{h,relax}$, indicating  that the system could not reach the steady state even at the end of the stroke. The time spent along the cold isochore is chosen  as   $\tau_{c}=5$s$\gg\tau_{c,relax}$ which indicates that $\rho_{t_0}$ is close to $\rho_{c}^{eq}$ . 
For the machine under consideration, 
the thermal contact  has the time of second scale, while the time spent on the driven stroke is always smaller than
$1$ ms, thereby indicating that the cycle period, $\tau_{cyc}=\tau_h+\tau_c+\tau_{ch}+\tau_{hc}$,   would be dominated by $\tau_h$ and $\tau_c$.

 \begin{figure*}
{\includegraphics[width=18cm]{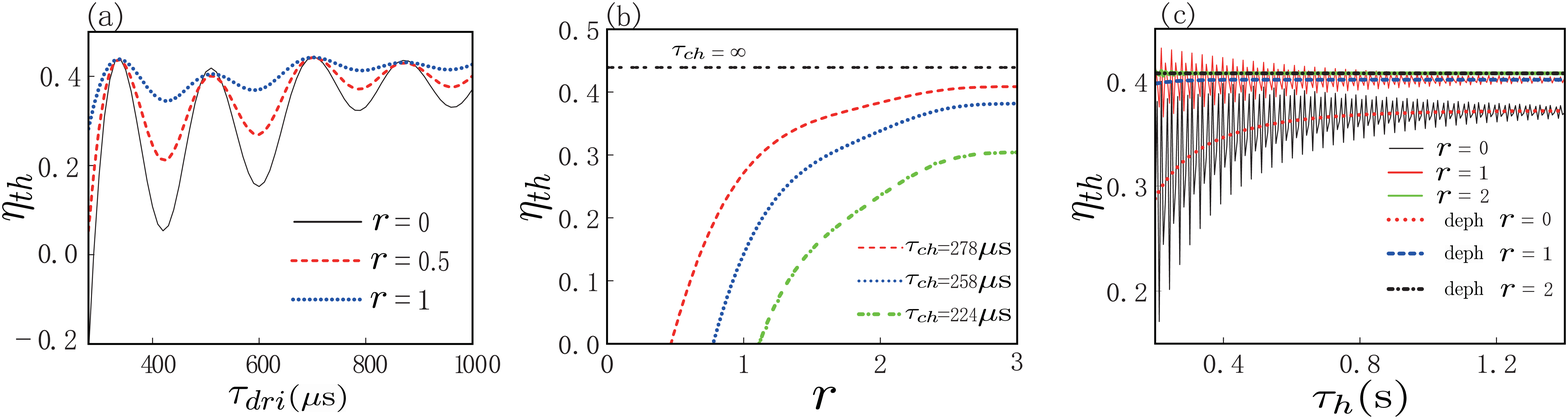}}
\caption{Thermodynamic efficiency $\eta_{th}$  as a function of  (a) driving time $\tau_{ch}$, (b) squeezing parameter $r$, and (c) thermal-contact time $\tau_h$. The parameters in (a), (b), and  (c) are same as   in  Figs.\ref{p} (a), \ref{p} (b), and \ref{p} (c), respectively.}
\label{eff}
\end{figure*}
\begin{figure*}[tb]
{\includegraphics[width=7.5cm]{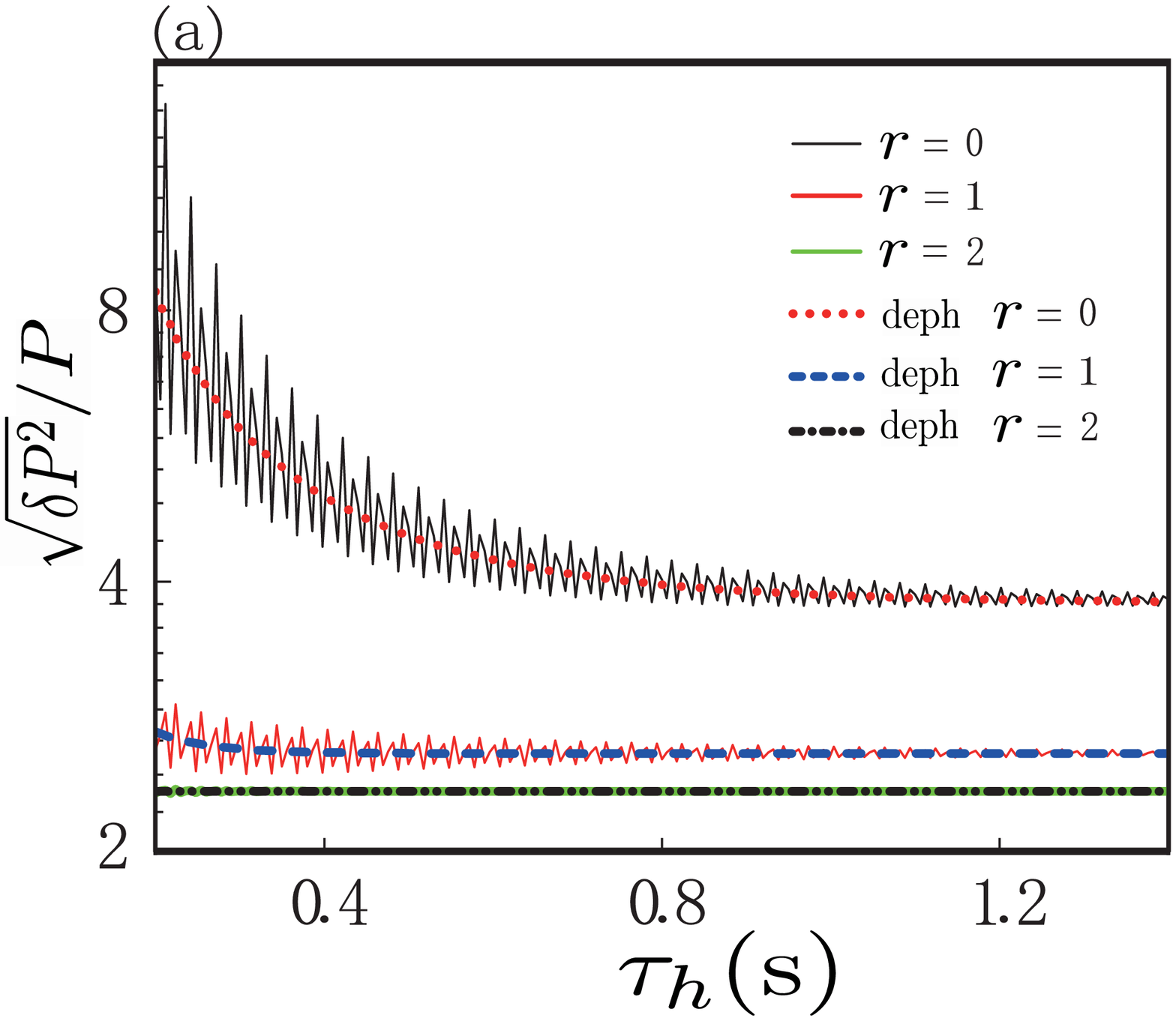}}
{\includegraphics[width=8.4cm]{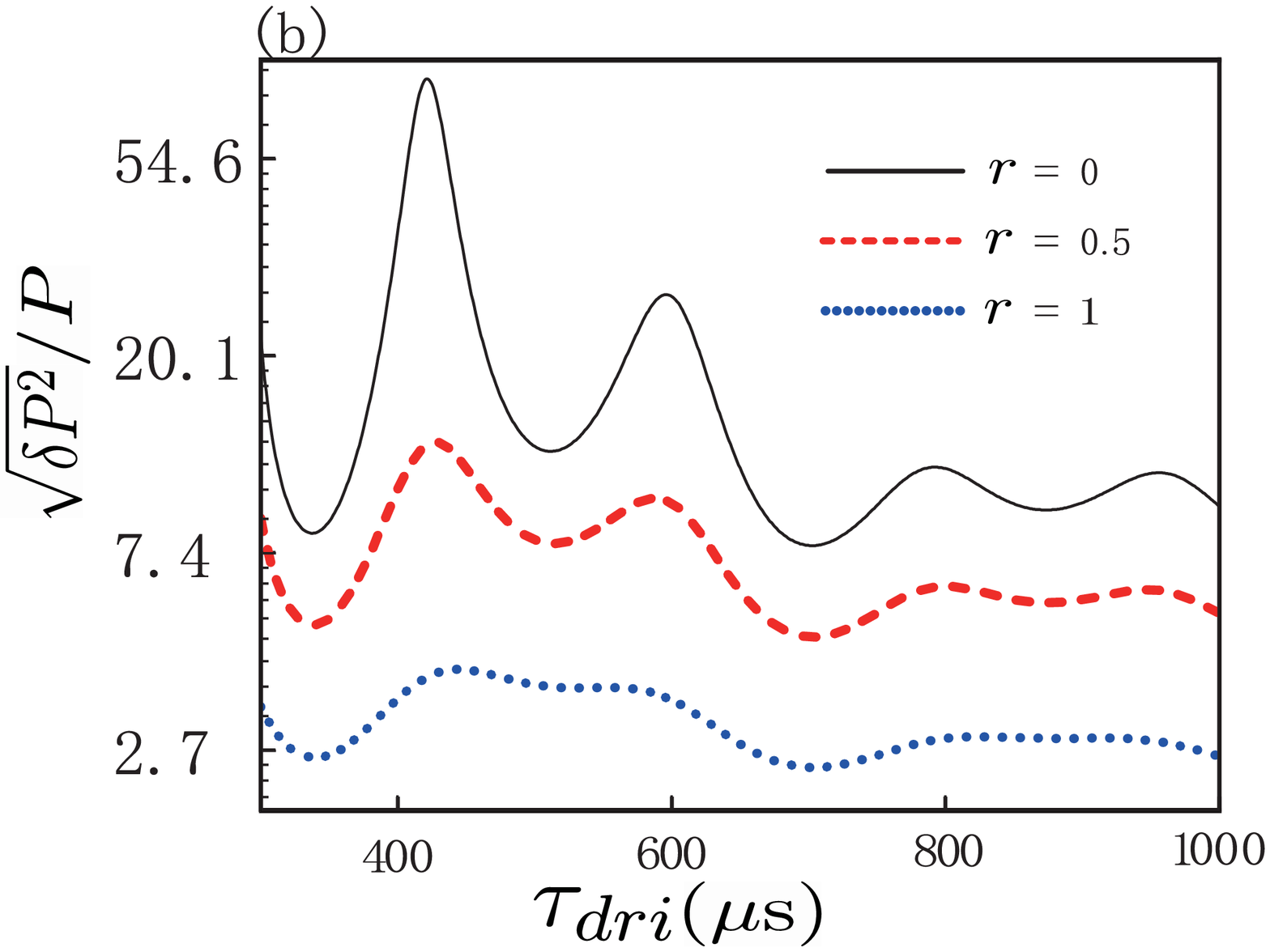}}
\caption{Root-mean-square relative fluctuation of   power, $\sqrt{\delta P^2}/P$, as a function of (a) thermal-contact time $\tau_h$ and (b) driving time $\tau_{dri}$.  A logarithmic scale is used in
the root-mean-square relative fluctuation (ordinate axis). The parameters in (a) and (b) are the same as in Fig. \ref{p}(a) and \ref{p}(b), respectively. }
\label{flup}
\end{figure*}
\begin{figure*}[tb]
{\includegraphics[width=9cm]{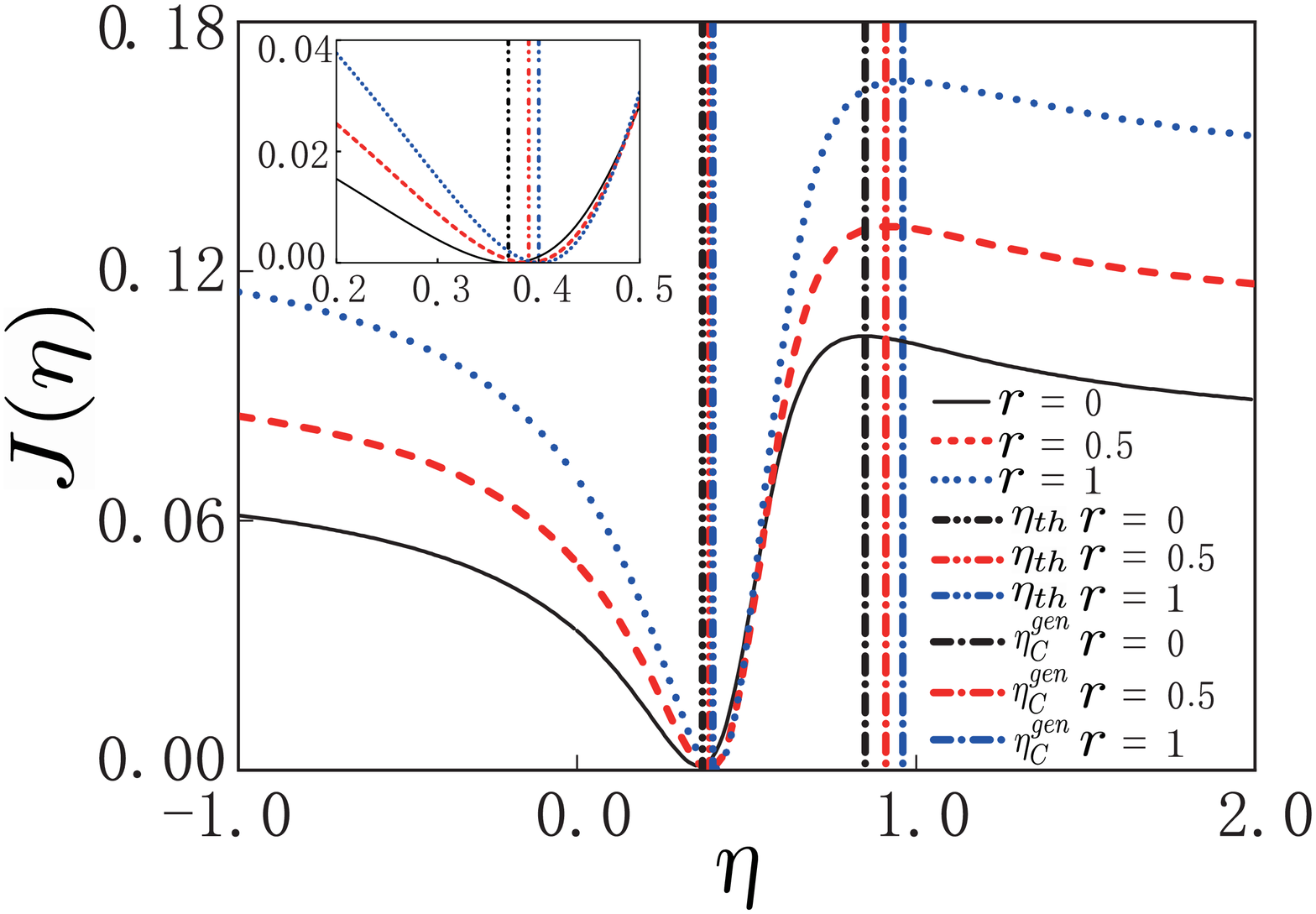}}
\caption{Large deviation function of  efficiency, $J(\eta)$, as a function of stochastic efficiency $\eta$. The parameters are $\tau_{dri}=460\mu$s, $\tau_h=2.8$s. $J(\eta)$ is situated between a maximum at the generalized Carnot efficiency $\eta_C^{gen}$ and a minimum at the thermodynamic efficiency $\eta_{th}$, covering the special case when squeezing was absent \cite{Ver14, Lut21}. Inset: Large deviation function $J(\eta)$ in the regime near the macroscopic efficiency. Both $\eta_{th}$   and $\eta_C^{gen}$ in Eq. (\ref{etage}) are increased with increasing $r$ as they should.}
\label{ldfe} 
\end{figure*}
In the limit of long driving time $\tau_{dri}$ , no coherence is generated in the first and third driven strokes, and thus the coherence $C(\rho_{t_3})$ should decay as $\tau_{ch}$ increases, as in Fig. \ref{cohd}(a), where the 
relative entropies of coherence for different values of squeezing parameter $r$  are qualitatively similar.  The coherence $C(\rho_{t_3})$ is always larger than $C(\rho_{t_2})$ [see Fig. \ref{cohd}(b)], and the difference between them is equivalent to the amount of coherence generated during the third stroke. Both Figs.  \ref{cohd}(a) and \ref{cohd}(b) show that the coherence is  a monotonic decreasing function of squeezing parameter $r$.  This can be understood by the fact that, in Eq. (\ref{rhot2}) the off-diagonal density matrix elements  (determined by $\mathcal{X}$ and $\mathcal{X}^*$ and related to the quantum coherence)  monotonically decrease as $r$ increases. 

Coherence indicates how far the system in contact with the squeezed bath deviates from the stationary  state. The squeezed reservoir reducing the coherence leads to a decrease in the LK divergence at the end of the hot isochore, as shown in Fig. \ref{lkd} (a).   The coherence generated in the first stroke   is only partially erased after the incomplete thermalization, and the residual  coherence at the instant  is thus decreasing as time $\tau_{dri}$ increases. Therefore, we observe from Fig. \ref{lkd}(a) that   $D(\rho_{t_2}||\rho_h^{ss})$   decreases as time $\tau_{ch}$.    Fig. \ref{lkd}(b) shows that, as expected,   $D(\rho_{t_2}||\rho_h^{ss})$  monotonically decreases as  $\tau_h$ for fixed $r$.      

Let us consider the average power which reads  $
    P={(\langle \mathrm{w}_{\mathrm{deph}}\rangle +\langle \mathrm{w}_{\mathrm{coh}}\rangle)}/\tau_{cyc}$, where  $\langle \mathrm{w}_{\mathrm{deph}}\rangle$ and $\langle \mathrm{w}_{\mathrm{coh}}\rangle$ were defined in Eqs. (\ref{wdep}) and (\ref{wcoh}), respectively. 
 In Figs. \ref{p}(a), \ref{p}(b) and \ref{p}(c), we plot the average power  $P$  as a function of driving time $\tau_{dri}$, squeezing parameter $r$, and thermal-contact time $\tau_h$, respectively. The power oscillates as a
function of the driving  time  $\tau_{dri}$, and very quick driving speed results in poor power output [Fig. \ref{p}(a)].  Under the condition that the total cycle period $\tau_{cyc}$ is dominated by $\tau_{h,c}$,
the major contribution of  time $\tau_{dri}$ associated with driving speed  to the power is contributed by quantum inner friction responsible for irreversible work along the two driven strokes. 

 Fig. \ref{p}(b) shows that the power $P$  increases monotonically  with increasing  squeezing parameter $r$. Increasing $r$ leads to a decrease in the so-called coherent work (\ref{wcoh}), and thus the total work (\ref{wtoh}) increases as $r$ increases. On the other hand, involving squeezing would shorten the total cycle period $\tau_{cyc}$  by decreasing  time $\tau_h$ for fixed divergence  $D(\rho_{t_2}||\rho_h^{ss})$ [Fig.\ref{lkd} (b)]. 
 
The thermal-contact-time variation  of the  power $P$ is sensitively dependent on the squeezing parameter $r$ (Fig. \ref{p}(c)]. We observe that the power first increases in small $\tau_h$ and then decreases with further increase in  $\tau_h$ for  $r=0$ or $r=1$, but the power decreases monotonically as $\tau_h$ increases for $r=2$.  
For small and vanishing squeezing ( $r=1,0$), fast hot isochoric stroke (relaxation process)  suppresses the decoherence
of the system. This implies that quick hot isochoric process requires additional,  coherent work $\langle \mathrm{w}_{\mathrm{coh}}\rangle$ compared to the slow isochoric evolution. $\langle \mathrm{w}_{\mathrm{coh}}\rangle$  decreases substantially as $\tau_h$ increases; the total work increases faster than linearly with increasing $\tau_h$. Therefore, the power increases with increasing $\tau_h$ to a certain maximum value, at which the effects of coherence are extremely small, and then decreases gradually.  
The oscillations in Fig. \ref{p}(c) comes from the effect of the dynamical interference  between
the residual coherence after the second stroke and the coherence generated in the
third stroke. Interestingly, in the large squeezing case ($r=2$), $\mathcal{X}$ in Eq. (\ref{rhot2}) is always negligible and $\langle \mathrm{w}_{tot}\rangle\rightarrow \langle \mathrm{w}_{\mathrm{deph}}\rangle$ ; hence, 
  zero coherence leads to  vanishing interference effect  and the total work  (\ref{wtoh}) becomes equivalent to  the work in dephased case.

Another benchmark parameter describing the machine performance is the thermodynamic efficiency $\eta$, in addition to the power $P$.
Efficiency  as a function of driving time $\tau_{dri}$ for different values of squeezing parameter is shown in Fig. \ref{eff}({a}), where the oscillation of $\eta$ comes from the inner friction captured by $\xi$.  The efficiency increases with increasing  driving time, although not monotonically. Fig. \ref{eff}(a) also shows that the oscillation amplitude decreases as $r$ increases, with the amplitude in $r=1$  much less than in $r=0$.   
It is again shown that the thermodynamic efficiency $\eta_{th}$ (\ref{etage}) increases with increasing $r$.  

As $r$ increases, the irreversible work induced by inner friction decreases. Therefore, increasing $r$  leads to an increase in efficiency [see also Fig. \ref{eff}(b)]  but a decrease in oscillation magnitude of efficiency. 
In  Fig. \ref{eff}(c), the efficiency 
corresponding to the   power  in Fig. \ref{p}(c) is shown. The
shapes of the efficiency  and power curves are similar, except that $\eta$ increases with $\tau_h$ to reach its maximum value consistent with Eq. (\ref{et11}). As emphasized, the effect of the dynamical interference on the thermodynamic efficiency behaves similarly to the power output [Fig. \ref{p}(c)].

 To describe the stability of the heat engine, we  consider the root-mean-square relative fluctuation of   power, $\sqrt{\delta P^2}/P$, which is equivalent to the coefficient of variation of the work, $\sqrt{\delta \mathrm{w}^2}/\langle \mathrm{w}\rangle$, and measures the dispersion of the probability distribution. Fig. \ref{flup}(a) and  \ref{flup}(b) display 
the square root of the relative power fluctuations, $\sqrt{ \delta \mathrm{w}^2}/\langle \mathrm{w}_{tot}\rangle$, as a function of thermal-contact  time $\tau_h$ and driving time $\tau_{dri},$ respectively.  Fig. \ref{flup}(a) shows that the oscillation timescale of power fluctuation with respect to the thermal-contact time $\tau_h$ agrees with the corresponding power [\ref{p}(c)] and efficiency [\ref{eff}(c)]. 
 However, the power fluctuation $\sqrt{\delta P^2}/P$ decreases monotonically while thermal-contact time  $\tau_h$ increases.  The larger the time of the system-bath interaction interval is, the closer the system to the stationary state, so the non-equilibrium thermal fluctuation of the power  is expected to decrease [see Fig. \ref{flup}(a)]. Increasing $r$ decreases the coherence generated in the unitary driven strokes, leading to decrease in the the magnitude of fluctuation oscillations. Note that,  for fast driving stroke,  the relative power fluctuation  is particularly large in the small and vanishing squeezing case, as demonstrated in Fig. \ref{flup}(b). This is because for fast driving speed, large irreversible work induced by both coherence and quantum friction results in small total work produced by the quantum engine.  However, under very large squeezing, the effect of  coherence is trivial [cf. Fig. \ref{p}(c)], leading to small relative power fluctuations. This implies that, the finite-time quantum Otto engines run more stably in the presence of squeezing than in the absence of squeezing.
 
 Since the average efficiency, $\langle\eta\rangle=\langle w/q_h\rangle$,  is ill-defined due to its divergence when the inner friction is present \cite{Lut20},  we investigate  
the efficiency statistics  in the long-time trajectories by determining the large deviation function of efficiency (\ref{jeta}). We plot the large deviation function of efficiency for  the quantum Otto engine in Fig. \ref{ldfe}, where the curve has a maximum when $\eta=\eta_C^{\mathrm{gen}}$ ($\eta_C^{\mathrm{gen}}$ reduces to $\eta_C$ if $r=0$) and a minimum at $\eta=\eta_{th}$. We therefore find that the standard thermodynamic
efficiency is the most likely value, and the generalized Carnot efficiency is the least likely. Furthermore, the
 rate function $J(\eta$) is  strictly larger in presence of squeezing  than the case without squeezing, with the exception of the point at
$\eta=\eta_{th}$. This shows that the convergence of the heat engine towards to the thermodynamic efficiency is improved by including the reservoir squeezing.


\section{conclusion}
Squeezed baths have been employed to fuel models of choice for investigating the novel performance of the quantum heat engines as compared to their macroscopic counterparts. Such machines proceeding finite time have been focused on the optimization on the average power at high temperature limit, with special emphasis on the efficiency at maximum power. 
 We have here extended these studies
to include the effects of  quantum coherence and inner friction, two essential quantum features.  In terms of squeezing parameter, we  have derived  compact
expressions of the power, thermodynamic efficiency, and relative power fluctuations for the quantum Otto engine.  We have found that the quantum Otto engine under squeezing can outperform its non-squeezing counterparts  by enhancing efficiency and power output with higher stability. 
We have additionally shown that squeezed reservoir leads to engines with better stability and faster convergence of efficiency to its most probable value. In addition, the numerical simulation adopted parameters  employed in
the recent experiment, which enable us to experience future experimental realization of high-performance quantum engines with good stability. 

\begin{acknowledgements}
This work was supported by the  NSFC  under Grants No. 11875034 and No. 61835013). W.-M. Liu also acknowledges the financial support by the National Key R\!$\!\And\!$\!D Program of China under Grants No. 2021YFA1400900, No.2021YFA0718300, and NO. 2021YFA1400243.
\end{acknowledgements}

\appendix

\section{The out power and heat of the cycle}\label{appower}
\numberwithin{equation}{section}
In this section, we derive the analytical expressions for averages of work and heat, and work fluctuations.
The joint distribution of quantum  work $ \mathrm{w}_{tot}$ and heat $q_h$ determines   the distribution functions of  work  and heat respectively.  For the two-level system, the ground and excited eigenenergies are  $\varepsilon_g^{\alpha}=-\hbar\omega_{\alpha}$ and $\varepsilon_e^{\alpha}=\hbar\omega_{\alpha}$, where  $\alpha=c, h$ correspond to the cold and hot isochores, respectively. It follows, using 
 Eq.(\ref{JD}), the probability distribution of work and heat can be calculated as 
\begin{widetext}
\begin{eqnarray}\label{pwd}
   p(\mathrm{w}_{tot})&=&\int dq_{h} p(\mathrm{w}_{tot},q_{h})\nonumber\\
   &=&[p_{t_0}^{e}p_{t_2}^{e}+p_{t_0}^{g}p_{t_2}^{g}-2(p_{t_0}^{e}p_{t_2}^{e}+p_{t_0}^{g}p_{t_2}^{g})\xi+\xi^{2}]\delta(\mathrm{w})\nonumber\\
   &+&2 \zeta_{hc} (\xi-p_{t_0}^{g})\delta(\mathrm{w}+\frac{\hbar w_{h}}{2})+p_{t_0}^{e}p_{t_2}^{e}\xi^{2}\delta(\mathrm{w}+\hbar w_{c}+\hbar w_{h})\nonumber\\
   &+&p_{t_0}^{g}(1-\xi)\xi\delta(\mathrm{w}-\hbar w_{c})+2p_{t_0}^{g}\zeta_{hc}(1-\xi)\delta(\mathrm{w}-\hbar w_{c}+\frac{\hbar w_{h}}{2})\nonumber\\
   &+&p_{t_2}^{e}(1-\xi)\xi\delta(\mathrm{w}+\hbar w_{h})+p_{t_0}^{g}p_{t_2}^{e}(1-\xi)^{2}\delta(\mathrm{w}-\hbar w_{c}+\hbar w_{h})\nonumber\\
   &+&2\zeta_{ch}(\xi-p_{t_2}^{g})\delta(\mathrm{w}+\frac{\hbar w_{c}}{2})+4\zeta_{ch}\zeta_{hc}\delta(\mathrm{w}+\frac{\hbar w_{c}+\hbar w_{h}}{2})\nonumber\\
   &+&2\zeta_{ch}(p_{t_2}^{e}-\xi)\delta(\mathrm{w}-\frac{\hbar w_{c}}{2})-4\zeta_{ch}\zeta_{hc}\delta(\mathrm{w}-\frac{\hbar w_{c}-\hbar w_{h}}{2}\nonumber)\\
   &-&2p_{t_2}^{e} \zeta_{ch}\xi\delta(\mathrm{w}+\frac{\hbar w_{c}}{2}+\hbar w_{h})+2p_{t_2}^{e} \zeta_{ch}(\xi-1)\delta(\mathrm{w}-\frac{\hbar w_{c}}{2}+\hbar w_{h})\nonumber\\
   &+&p_{t_2}^{g}(1-\xi)\xi\delta(\mathrm{w}-\hbar w_{h})+2\zeta_{hc}(p_{t_0}^{e}-\xi)\delta(\mathrm{w}-\frac{\hbar w_{h}}{2})\nonumber\\
   &+&p_{t_0}^{g} p_{t_2}^{g} \xi^{2}\delta(\mathrm{w}-\hbar w_{c}-\hbar w_{\tau_{h}})+2p_{t_0}^{g} \zeta_{hc}\xi\delta(\mathrm{w}-\hbar w_{c}-\frac{\hbar w_{h}}{2})\nonumber\\
   &+&2p_{t_2}^{g} \zeta_{ch}(1-\xi)\delta(\mathrm{w}-\hbar w_{h}+\frac{\hbar w_{c}}{2})-4\zeta_{ch} \zeta_{hc}\delta(\mathrm{w}-\frac{\hbar w_{h}-\hbar w_{c}}{2})\nonumber\\
   &+&2p_{t_2}^{g} \zeta_{ch}\xi\delta(\mathrm{w}-\frac{\hbar w_{c}}{2}-\hbar w_{h})+4 \zeta_{ch} \zeta_{hc}\delta(\mathrm{w}-\frac{\hbar w_{c}+\hbar w_{h}}{2})\nonumber\\
   &+&p_{t_0}^{e}(1-\xi)\xi\delta(\mathrm{w}+\hbar w_{c})-2p_{t_0}^{e}\zeta_{hc} \xi\delta(\mathrm{w}+\hbar w_{c}+\frac{\hbar w_{h}}{2})\nonumber\\
   &+&p_{t_0}^{e} p_{t_2}^{g}(1-\xi)^2\delta(\mathrm{w}-\hbar w_{h}+\hbar w_{c})+2p_{t_0}^{e}\zeta_{hc}(\xi-1)\delta(\mathrm{w}-\frac{\hbar w_{h}}{2}+\hbar w_{c}),
 \end{eqnarray}
 and
 \begin{eqnarray}\label{pqh}
 p(q_{h})&=&\int d\mathrm{w}p(\mathrm{w},q_{h})\nonumber\\
 &=&\{[p_{t_0}^{g}(1-\xi)+p_{t_0}^{e}\xi-2\zeta_{ch}]p_{t_2}^{g}+[p_{t_0}^{e}(1-\xi)+p_{t_0}^{g}\xi+2\zeta_{ch}]p_{t_2}^{e}\}\delta(q_{h})\nonumber\\
 &+&[p_{t_0}^{g}(1-\xi)+p_{t_0}^{e}\xi-2\zeta_{ch}]p_{t_2}^{e}\delta(q_{h}-\hbar w_{h})
 +[p_{1}^{e}(1-\xi)+p_{t_0}^{g}\xi+2\zeta_{ch}]p_{t_2}^{g}\delta(q_{h}+\hbar w_{h}).
 \end{eqnarray}
Here and hereafter we use $p_{t_0}^{g}=\langle g(t_0)|\rho_{t_0}|g(t_0)\rangle$, $p_{t_0}^{e}=\langle e(t_0)|\rho_{t_0}|e(t_0)\rangle$, $p_{t_2}^{g}=\langle g(t_2)|\rho_{t_2}|g(t_2)\rangle$, $p_{t_2}^{e}=\langle e(t_2)|\rho_{t_2}|e(t_2)\rangle$, $\zeta_{ch}=-Re[U_{ch}^{gg}\langle g(t_0)|\rho_{t_0}|e(t_0)\rangle U_{ch}^{eg\dag}]$, $\zeta_{hc}=-Re[U_{hc}^{gg}\langle g(t_2)|\rho_{t_2}|e(t_2)\rangle U_{hc}^{eg\dag}]$.
Integrating over the probability distribution function Eq. (\ref{pwd}) allows us to find the expressions for the first two central moments of quantum work, which are
\begin{eqnarray}\label{wdw}
-\langle \mathrm{w}\rangle&=&-\int d\mathrm{w}p(\mathrm{w})\mathrm{w}\nonumber\\
&=&\hbar(w_{h}-w_{c})(p_{t_0}^{g}p_{t_2}^{e}-p_{t_0}^{e}p_{t_2}^{g})-2\hbar w_{h}\zeta_{ch}-2\hbar w_{c}\zeta_{hc}\nonumber\\
&-&[\hbar (w_{c}-w_{h})(p_{t_0}^{e}p_{t_2}^{g}-p_{t_0}^{g}p_{t_2}^{e})+\hbar (w_{0}+w_{h})(p_{t_0}^{g}p_{t_2}^{g}-p_{t_0}^{e}p_{t_2}^{e})]\xi\nonumber\\
&=&\hbar(w_{h}-w_{c})(\langle n_{t_2}\rangle-\langle n_{t_0}\rangle)
+2\hbar\xi(w_{c}\langle n_{t_2}\rangle+w_{h}\langle n_{t_0}\rangle)\nonumber\\
&-&
2\hbar w_{h}\zeta_{ch}-2\hbar w_{c}\zeta_{hc},
\end{eqnarray}
and
\begin{eqnarray}\label{w2dw}
\langle\mathrm{w}^{2}\rangle&=&\int d\mathrm{w}p(\mathrm{w})\mathrm{w}^{2}\nonumber\\
&=&\hbar^2 w_c^2[p_{t_0}^g p_{t_2}^e+p_{t_0}^e p_{t_2}^g +(p_{t_0}^e-p_{t_0}^g)(p_{t_2}^e-p_{t_2}^g)\xi-2(p_{t_0}^e-p_{t_0}^g)\zeta_{hc}]\nonumber\\
&+&\hbar^2 w_{h}^{2}[p_{t_0}^g p_{t_2}^e+p_{t_0}^e p_{t_2}^g +(p_{t_0}^e-p_{t_0}^g)(p_{t_2}^e-p_{t_2}^g)\xi-2(p_{t_2}^e-p_{t_2}^g)\zeta_{ch}]\nonumber\\
&+&\hbar^2w_h w_c[2(p_{t_0}^g p_{t_2}^e+p_{t_0}^e p_{t_2}^g)(2\xi-1)+2(p_{t_0}^e-p_{t_0}^g)(p_{t_2}^e-p_{t_2}^g)\xi^2\nonumber\\
&+&2\zeta_{hc}(p_{t_0}^e-p_{t_0}^g)+2\zeta_{ch}(p_{t_2}^e-p_{t_2}^g)-4\zeta_{hc}\xi(p_{t_0}^e-p_{t_0}^g)-4\zeta_{ch}\xi(p_{t_2}^e-p_{t_2}^g)+8\zeta_{ch}\zeta_{hc}]\nonumber\\
&=&w_h^2[1/2 - 2 \langle n_{t_2}\rangle (\langle n_{t_0}\rangle + 2 \zeta_{ch} - 2 \langle n_{t_0}\rangle \xi)]+w_c^2 [1/2 - 2 \langle n_{t_0}\rangle (\langle n_{t_2}\rangle+ 2 \zeta_{hc} - 2 \langle n_{t_2}\rangle \xi)]\nonumber\\
&+&w_c w_h[2 \xi - 1 + 4 \zeta_{ch} \langle n_{t_2}\rangle (1 - 2 \xi) + 4\zeta_{hc} \langle n_{t_0}\rangle(1 - 2\xi) + 
 4 \langle n_{t_0}\rangle \langle n_{t_2}\rangle (2 \xi^2 + 1 - 2 \xi) + 8 \zeta_{ch} \zeta_{hc}].
\end{eqnarray}
In deriving Eqs.(\ref{wdw}) and (\ref{w2dw}), we have used,  $\langle n_{t_0}\rangle=({p_{t_0}^e-p_{t_0}^g})/2$ and $\langle n_{t_2}\rangle=({p_{t_2}^e-p_{t_2}^g})/{2}$, which are average populations of the system at time $t=t_0$ and $t=t_2$, respectively. By combining these two equations, we can obtain the work fluctuations, $\delta \mathrm{w}^2=\langle \mathrm{w}^2\rangle-\langle \mathrm{w}\rangle^2$ [cf. Eq.(\ref{wfluc}) in the main text].
By very simple algebra, we use  Eq. (\ref{pqh}) to obtain the macroscopic heat injection, $\langle q_h\rangle=\int dq_{h}q_{h}p(q_{h})
 =\hbar w_{h}[p_{t_0}^{g}p_{t_2}^{e}-p_{t_0}^{e}p_{t_2}^{g}+(p_{t_0}^{e}-p_{t_0}^{g})\xi-2\zeta_{ch}]$, which, in terms of the populations $n_{t_0}$ and $n_{t_2}$, can be re-expressed by Eq. (\ref{qh}). 
\end{widetext}

\section{Efficiency and Entropy production}\label{apeta}
  When the open quantum system evolves along an isochoric process,   the external field is frozen and the control parameter is fixed, leading to static system Hamiltonian $H(t)=H$.  The dynamics of the system with density operator $\rho_t$ is well described 
 by the master equation of
Lindblad form \cite{MM97, HF02, Kos14, Wang19}:
\begin{equation}
\frac{{d\rho_t}}{dt}=-\frac{i}{\hbar}[{H},{\rho_t}]+\mathcal{L}_D({\rho_t}),
\label{drhoa}
\end{equation}
 where $\mathcal{L}_D({\rho})=\sum_\alpha {L}_\alpha\rho
{L}_\alpha^\dag-\frac{1}2\left[{L}_\alpha^\dag
{L}_\alpha,{\rho}\right]_+$, with $[~~]_+$ being  the anticommutator. The dissipation term 
$\mathcal{L}_D(\rho)$  is
responsible for driving the quantum system to  relax to a unique, positive-definite, steady state with density matrix $\rho^{eq}_t$, given as the solution of $\mathcal{L}_D (\rho^{eq}_t) = 0$. In the two-level system, the
  Kraus operators ${L}_\alpha$ can be written  in terms  of the quantum jump operators, namely,  $L_-=\sigma_-\sqrt{N^{th}+1}$ and $L_+=\sigma_+\sqrt{N^{th}}$,  with $N^{th}=1/(e^{\beta\omega}-1)$,  and then Eq.(\ref{drhoa}) reduces to the form given by Eq. (\ref{drhoc}) in the main part.  The system under consideration would reach equilibrium with the thermal bath
if the system-bath interaction time is long enough. Such
a Gibbs thermal state for the system can be obtained by determining the steady state solution of differential equation (\ref{drhoa}) to arrive at 
  \begin{equation}
\rho^{eq}_t=\rho^{eq}=\frac{e^{-\beta H}}{\mathrm{Tr}\left(e^{-\beta H}\right)}.\label{reqa}
 \end{equation}

  Let us  consider the case of the two-level system interacting with a squeezed thermal bath of inverse temperature $\beta$.  We introduce the Lindblad operators \cite{Kos14, Manz16}
$\mathcal{S}_-=\sigma_-\cosh(r)+\sigma_+\sinh(r) e^{i\theta}$ and $\mathcal{S}_+=\sigma_+\cosh(r)+\sigma_-\sinh(r) e^{-i\theta}$ to write the Kraus operators as $L_-^{sq}=\sqrt{N^{th}+1}\mathcal{S}_-$ and $L_+^{sq}=\mathcal{S}_+\sqrt{N^{th}}$.  Equation (\ref{drho}) in the main text can then  be re-expressed   in the Lindblad form as
\begin{equation} \label{drhos}
\frac{{d\rho_t}}{dt}=\mathcal{L}^{sq}_D(\rho_t)=\sum_{\alpha\neq\alpha'}  {L}_\alpha^{sq}\rho_t
{L}_{\alpha'}^{sq}-\frac{1}2\left[{L}_{\alpha'}^{sq}
{L}_\alpha^{sq},{\rho_t}\right]_+,
\end{equation}
with $\alpha=-,+$.
As we show in the main part, we can introduce the effective inverse temperature of the squeezed bath,
$\beta^{\mathrm{eff}}=\frac{1}{\hbar\omega}\ln\frac{N^{ss}+1}{N^{ss}}$, where  the excitation number can be written in terms of the squeezing parameter $r$,
   $
   N^{ss}= N^{th}+(2 N^{th}+1)\sinh^2(r).$   
Under squeezing,  these  Lindblad operators $L_{\pm}^{sq}$ in Eq. (\ref{drhos})  satisfy the detailed balance
condition, ensuring that  the system  with frequency $\omega_t=\mathrm{const}$ can evolve asymptotically in
a specific way to the unique, steady  state  $\rho_t^{ss}$, namely, $\mathcal{L}_D^{sq}(\rho_t^{ss})=0$. The steady-state solution of Eq. (\ref{drhos}) for the open system of Hamiltonian $H(t)$ can be easily obtained, 
 \begin{equation}
  \rho^{ss}_t=\rho^{ss}=\frac{e^{-\beta^{\mathrm{eff}}H(t)}}{\mathrm{Tr}\left(e^{-\beta^{\mathrm{eff}}H(t)}\right)}. \label{rssa}
  \end{equation}

  For  a Markovian open quantum system, the irreversiblity  can be well described by the so-called nonadiabatic  entropy production  rate \cite{Par13,Bro10, Manz16},
\begin{equation}
  \Dot{\Sigma}=-\frac{d}{dt}D(\rho_t||\rho^{ss}_t)=\dot{S}_t-\dot{\Psi}_t,
\end{equation}
where $D(\rho_t||\rho^{ss}_t)=\mathrm{Tr}[\rho_t(\ln\rho_t-\ln\rho^{ss}_t)]$
is the quantum relative entropy. Here $\dot{S}_t$ can be given by $\dot{S}_t=-\mathrm{Tr} (\dot{\rho}_t\ln\rho_t+\dot{\rho}_t)=-\mathrm{Tr} (\dot{\rho}_t\ln\rho_t)$, and it indicates the rate of  change of the system's von Neumann entropy $S_t=-\mathrm{Tr}(\rho_t\ln \rho_t)$. The second contribution $\dot{\Psi}$, called the excess entropy production rate, is determined by 
\begin{equation}
\dot{\Psi}=-\mathrm{Tr}(\dot{\rho}_t\ln\rho^{eq}_t), \label{psi}
\end{equation} 
 which defines the effective rate
of entropy flow into the system from its surroundings. Under squeezing, the excess entropy production rate can be expressed as 
\begin{equation}
\dot{\Psi}^{sq}=-\mathrm{Tr}(\dot{\rho}_t\ln\rho_t^{ss}). \label{psisq}
\end{equation}

 We are now in a position to discuss the entropy production for the quantum Otto engine sketched in Fig. \ref{model}. The machine consists of two isentropic (unitary) compression, and two isochoric heating stroke  where $H(t)=H_h$ and  cooling stroke  with $H(t)=H_c$.  The working substance returns to its initial state after a single cycle and thus its entropy change is zero, which means $\int_0^{\tau_{cyc}}\dot{S}(t)dt=0$, $\tau_{cyc}=\tau_h+\tau_c+2\tau_{dri}$. We  therefore find that, for the cyclic engine,  the nonadiabatic entropy $\Sigma$ is    equivalent to the excess entropy and it is coming exclusively from the two system-bath interaction intervals. As emphasized, by incorporating the condition of detailed balance in the Lindblad master equations [c.f. Eq. (\ref{drhos}) or  (\ref{drhoa})], the working system  in the long time intervals of system-bath interaction should reach the thermal state  $\rho^{eq}$ in the absence of reservoir squeezing  or steady state $\rho^{ss}$ in the presence of  squeezing.
 For the engine cycle, the total entropy production turns out to be   $\Sigma_{tot}=\int_{0}^{\tau_{cyc}}\dot{\Sigma}dt=-\int_{t_1}^{t_2}\dot{\Psi}^{sq}dt-\int_{t_3}^{\tau_{cyc}}\dot{\Psi}^{sq}dt$, where $\dot{\Psi}^{sq}$ was defined by  Eq . (\ref{psisq}).
 
 Hence, the irreversible production $\Sigma_{tot}$ of the cycle can  be calculated as 
$\Sigma_{tot}
  =-[\mathrm{Tr}(\rho_{t_2}\ln\rho_h^{ss})-\mathrm{Tr}(\rho_{t_1}\ln\rho_h^{ss}]-[\mathrm{Tr}(\rho_{t_0}\ln\rho_c^{eq})-\mathrm{Tr}(\rho_{t_3}\ln\rho_c^{eq}]$, where $\rho_h^{ss}$  can be determined according to  Eq. (\ref{rssa}) by  setting $\beta^{\mathrm{eff}}=\beta_h^\mathrm{{eff}}$ and $H(t)=H_h$, and $\rho_c^{eq}$ according to   Eq. (\ref{reqa}) by using $\beta=\beta_c$ and $H(t)=H_c$. It then follows that the total entropy production is given by 
    \begin{eqnarray}\label{Stot}
    \Sigma_{tot}&=&
    -\beta_h^{\mathrm{eff}}[\mathrm{Tr}(\rho_{t_2}H_{h})]-\mathrm{Tr}(\rho_{t_1}H_{h})]\nonumber\\
  &-&\beta_{c}[\mathrm{Tr}(\rho_{t_0}H_c)-\mathrm{Tr}(\rho_{t_3}H_c)]\nonumber\\
  &=&-\beta_{h}^{\mathrm{eff}}\langle q_{h}\rangle-\beta_{c}\langle q_{c}\rangle,
\end{eqnarray} 
where we have used $\langle q_h\rangle=\mathrm{Tr}(\rho_{t_2}H_{h})-\mathrm{Tr}(\rho_{t_1}H_{h})$ and $\langle q_c\rangle=\mathrm{Tr}(\rho_{t_0}H_c)-\mathrm{Tr}(\rho_{t_3}H_c)$. 
Using Eq.(\ref{Stot}), we derive the machine efficiency as 
\begin{equation}
  \eta=1+\frac{\langle q_{c}\rangle}{\langle q_{h}\rangle}
  =1-\frac{\beta_{h}^{\mathrm{eff}}\langle q_{h}\rangle+\Sigma_{tot}}{\beta_c\langle q_{h}\rangle}.
\end{equation}
Introducing the so-called generalized efficiency  $\eta_{C}^{gen}=1-{\beta_{h}^{\mathrm{eff}}}/{\beta_{c}}$, we then obtain Eq. (\ref{etage}) in the main text.

\end{document}